\documentclass[a4paper,11pt]{article}

\pdfoutput=1  

\usepackage{amsmath,amssymb,mathtools}     
\usepackage{color}
\usepackage{graphicx}
\usepackage{subfigure}
\usepackage{cite}                
\usepackage{hyperref}            
\usepackage{multirow,makecell}   
\usepackage{textcomp}
\usepackage{wasysym}

\usepackage[text={16.3cm,24cm},centering]{geometry}

\numberwithin{equation}{section}   

\def \be {\begin{equation}}
\def \ee {\end{equation}}
\def \ba {\begin{array}}
\def \ea {\end{array}}
\def \bea{\begin{eqnarray}}
\def \eea{\end{eqnarray}}
\def \nn {\nonumber}

\def \a {\alpha}
\def \b {\beta}
\def \g {\gamma}

\def \G {\Gamma}
\def \d {\delta}
\def \D {\Delta}
\def \e {\epsilon}

\def \s {\sigma}

\def \r {\rho}

\def \t {\tau}

\def \mA {\mathcal A}

\def \mC {\mathcal C}

\def \mF {\mathcal F}

\def \mR {\mathcal R}
\def \mS {\mathcal S}

\def \mX {\mathcal X}
\def \mY {\mathcal Y}
\def \mZ {\mathcal Z}

\def \p {\partial}
\def \f {\frac}
\def \df {\dfrac}

\def \lt {\left}
\def \rt {\right}

\def \td {\tilde}
\def \hs {\hspace}

\def \inf {\infty}

\def \lag {\langle}
\def \rag {\rangle}

\def \ep {\mathrm{e}}
\def \ii {\mathrm{i}}

\def \Re {{\textrm{Re}}}
\def \Im {{\textrm{Im}}}
\def \tr {\textrm{tr}}

\def \and {{\textrm{and}}}

\def \for {{\textrm{for}}}
\def \when {{\textrm{when}}}

\def \pl {{\textrm{pl}}}
\def \gr {{\textrm{gr}}}
\def \cft {{\textrm{cft}}}

\def \sh {{\textrm{sh}}}
\def \lo {{\textrm{lo}}}
\def \thm {{\textrm{th}}}

\def \CFT {{\textrm{CFT}}}

\def \I {{\textrm{I}}}
\def \II {{\textrm{II}}}
\def \III {{\textrm{III}}}
\def \IV {{\textrm{IV}}}
\def \V {{\textrm{V}}}

\def \cL {{\textrm{L}}}
\def \NL {{\textrm{NL}}}

\begin{document}

\title{\textbf{Corrections to holographic entanglement plateau}}
\author{Bin Chen$^{1,2,3}$\footnote{bchen01@pku.edu.cn}~,
Zhibin Li$^{4,5,6}$\footnote{lizb@ihep.ac.cn}~
and
Jia-ju Zhang$^{7,8}$\footnote{jiaju.zhang@mib.infn.it}}
\date{}

\maketitle

\vspace{-10mm}

\begin{center}
{\it
$^1$Department of Physics and State Key Laboratory of Nuclear Physics and Technology,\\Peking University, 5 Yiheyuan Road, Beijing 100871, China\\\vspace{1mm}

$^2$Collaborative Innovation Center of Quantum Matter, 5 Yiheyuan Road, Beijing 100871, China\\\vspace{1mm}

$^3$Center for High Energy Physics, Peking University, 5 Yiheyuan Road, Beijing 100871, China\\\vspace{1mm}

$^4$Theoretical Physics Division, Institute of High Energy Physics, Chinese Academy of Sciences,\\
19B Yuquan Road, Beijing 100049, China\\ \vspace{1mm}

$^5$Theoretical Physics Center for Science Facilities, Chinese Academy of Sciences,\\19B Yuquan Road, Beijing 100049, China\\ \vspace{1mm}

$^6$School of Physics Sciences, University of Chinese Academy of Sciences,\\19A Yuquan Road, Beijing 100039, China\\\vspace{1mm}

$^7$Dipartimento di Fisica, Universit\`a degli Studi di Milano-Bicocca,\\Piazza della Scienza 3, I-20126 Milano, Italy\\\vspace{1mm}

$^8$INFN, Sezione di Milano-Bicocca, Piazza della Scienza 3, I-20126 Milano, Italy
}
\end{center}

\vspace{8mm}

\begin{abstract}

  We investigate the robustness of the Araki-Lieb inequality in a two-dimensional (2D) conformal field theory (CFT) on torus. The inequality requires that $\Delta S=S(L)-|S(L-\ell)-S(\ell)|$ is nonnegative, where $S(L)$ is the thermal entropy and $S(L-\ell)$, $S(\ell)$ are the entanglement entropies. Holographically there is an entanglement plateau in the BTZ black hole background, which means that there exists a critical length such that when $\ell \leq \ell_c$ the inequality saturates $\Delta S=0$. In thermal AdS background, the holographic entanglement entropy leads to $\Delta S=0$ for arbitrary $\ell$. We compute the next-to-leading order contributions to $\Delta S$ in the large central charge CFT at both high and low temperatures. In both cases we show that $\Delta S$ is strictly positive except for $\ell = 0$ or $\ell = L$. This turns out to be true for any 2D CFT. In calculating the single interval entanglement entropy in a thermal state, we develop new techniques to simplify the computation. At a high temperature, we ignore the finite size correction such that the problem is related to the entanglement entropy of double intervals on a complex plane. As a result, we show that the leading contribution from a primary module takes a universal form. At a low temperature, we show that the leading thermal correction to the entanglement entropy from a primary module does not take a universal form, depending on  the details of the theory.

\end{abstract}

\baselineskip 18pt
\thispagestyle{empty}
\newpage



\tableofcontents

\section{Introduction}

The holographic entanglement entropy \cite{Ryu:2006bv,Ryu:2006ef,Nishioka:2009un} relates the quantum gravity to quantum information, and opens a new window to study the AdS/CFT correspondence \cite{Maldacena:1997re,Gubser:1998bc,Witten:1998qj,Aharony:1999ti}. The entanglement entropy in a quantum field theory is usually not easy to compute. For a conformal field theory (CFT) dual to the AdS Einstein gravity, it was suggested in \cite{Ryu:2006bv,Ryu:2006ef} that the entanglement entropy of a subregion $A$ could be holographically computed by the so-called Ryu-Takayanagi (RT) formula
 \be
 S_A =\frac{\mbox{Area of }\gamma_A}{4G_N},
 \ee
 where $\gamma_A$ is the minimal surface in the bulk homologous to the subregion $A$.
 The area law of the RT formula indicates a deep relation between the holographic entanglement entropy and the black hole entropy. It has actually been shown in \cite{Lewkowycz:2013nqa} that the holographic entanglement entropy is actually a kind of generalized gravitational entropy. More precisely, the RT formula originates from the semi-classical Euclidean gravity action, and there could be gravitational quantum corrections to the holographic entanglement entropy \cite{Fujita:2009kw,Headrick:2010zt,Barrella:2013wja,Faulkner:2013ana}. 

One of the situations that the quantum corrections to the holographic entanglement entropy are important is the so-called holographic entanglement plateau \cite{Hubeny:2013gta}. For a subsystem $A$ and its complement $A^c$ in a thermal state the Araki-Lieb inequality \cite{Araki:1970ba} requires that
\be
\D S=S_\thm - | S_{A^c} - S_A | \geq 0, \label{AL}
\ee
with $S_\thm$ being the thermal entropy of the whole system and $S_A$, $S_{A^c}$ being the entanglement entropies. For the holographic entanglement entropies, when the subsystem $A$ is small enough but still finite the inequality could be saturated at a high enough temperature \cite{Ryu:2006bv,Headrick:2007km,Blanco:2013joa,Hubeny:2013gta}. The saturation is called the entropy plateaux. In this case, the minimal surface $\g_{A^c}$ for the region $A^c$ is the disconnected sum of the minimal surface $\g_A$ for the region $A$ and the horizon of the black hole corresponding to the thermal state. However, the saturation is possible if only the classical contribution has been considered. It was pointed out in \cite{Faulkner:2013ana} that quantum corrections to the holographic entanglement entropy can resolve the saturation. In other words, after considering the quantum correction, there is always $\D S>0$, except for the case that the size of $A$ or $A^c$ becomes vanishing.

From the AdS/CFT correspondence, the classical action of the bulk configuration corresponds to the leading order contribution in the field theory at large $c$ (or $N$) limit, while the one-loop quantum correction corresponds to the next-to-leading order contribution. Such quantum correction is usually hard to compute in the bulk side\cite{Faulkner:2013ana}. In the case of AdS$_3$/CFT$_2$\cite{Brown:1986nw,Strominger:1997eq}, one may find the gravitational configuration via the Schottky uniformization\cite{Hartman:2013mia,Faulkner:2013yia} and compute the one-loop corrections by using the heat kernel and the image method\cite{Giombi:2008vd,Chen:2015uga}. However, in the large interval limit at finite temperature, the computation becomes complicated and needs appropriate treatment on the monodromy condition\cite{Chen:2015kua}. On the other hand, the large interval limit is singular in the sense that the usual  level expansion of the thermal density matrix becomes ill-behaved under the limit. One has to find another kind of  expansions to get the partition function perturbatively. In \cite{Chen:2014hta}, it was proposed that one has to insert the complete set of basis of the twist sector to compute the partition function. For the large interval at a high temperature, this proposal gives consistent results for the large $c$ CFT with the bulk computation\cite{Chen:2015kua}.

In this paper we revisit the issue of the large interval entanglement entropy and pay special attention to the corrections to the entanglement plateau\footnote{Strictly speaking, the entanglement plateau only appear in the high temperature CFT with large central charges. Here we refer to the quantity $\D S$ loosely as the entanglement plateau even in the low temperature case. The CFT at the low temperature is dual to the thermal AdS background, and the holographic entanglement entropy trivially leads to $\Delta S=0$ for arbitrary size of $A$.} in AdS$_3$/CFT$_2$. We mainly work on two-dimensional large central charge CFT with a sparse light spectrum\cite{Hartman:2013mia,Hartman:2014oaa}, which is dual to the semiclassical limit of AdS$_3$ gravity. On the CFT side we first focus on the vacuum module in the large $c$ limit, compute the short interval and long interval expansions of the entanglement entropies, and get nonvanishing corrections to the entanglement plateau. Moreover we also consider the leading contribution from a primary module. The contribution is at the next-to-leading order in the large $c$ limit. We find that in the high temperature case the correction from the  primary operator takes an universal form, but in the low temperature case the correction is not universal and takes a complicated form.

Though we mainly do computation in the large $c$ CFT, the study can actually be applied to a general 2D CFT as well. In a 2D CFT, the vacuum module plays an essential role as it involves the stress  tensor and its contribution to the entanglement entropy includes the part proportional to the central charge. Most of the study in this paper can be used in a general 2D CFT. The only thing one should be cautious is the large $c$ expansion, which could not make sense.

The rest of the paper is organized as follows.
In section~\ref{sec2} after giving a brief review of the holographic entanglement plateau,  we investigate $\D S$ (\ref{AL}) in the high temperature case. We show that after omitting the finite size correction, which is exponentially small in the high temperature limit, we can relate the computation to the one for the two-interval entanglement entropy on a complex plane. Therefore we are allowed to read the mutual information and the universal correction from a nonvacuum module.
In section~\ref{sec3}, we discuss the low temperature case with contributions from only the vacuum module using the method of multi-point correlation functions.
We conclude in section~\ref{sec5} with discussions.
In appendix~\ref{appA} we review the mutual information of two intervals on a complex plane that is useful for section~\ref{sec2}.
In appendix~\ref{appB} we calculate the relation relation (\ref{z8}) that is useful to sections~\ref{sec3}.
In appendix~\ref{appC}, we apply the operator product expansion (OPE) of the twist operators to compute for the low temperature case and find agreement with the results in sections~\ref{sec3}.

\section{High temperature case} \label{sec2}



\begin{figure}[htbp]
  \centering
  \subfigure[]{\includegraphics[height=35mm]{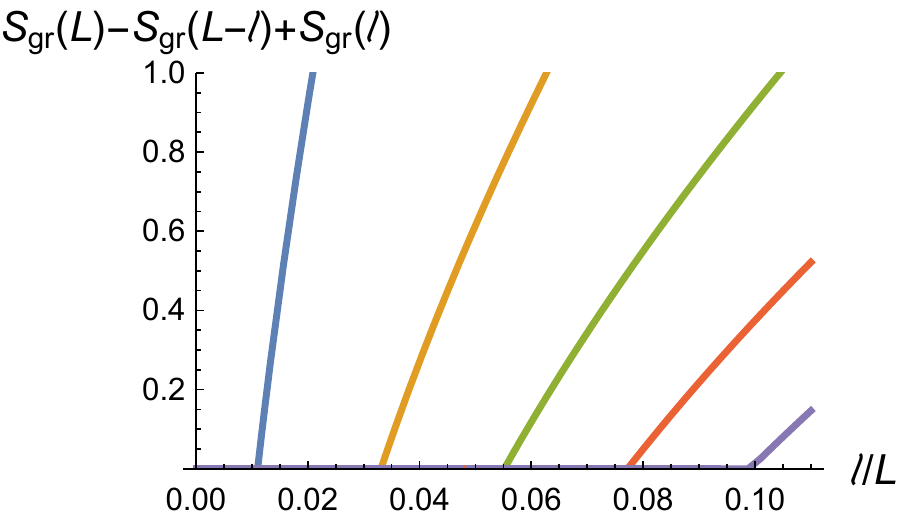}} ~~~
  \subfigure[]{\includegraphics[height=35mm]{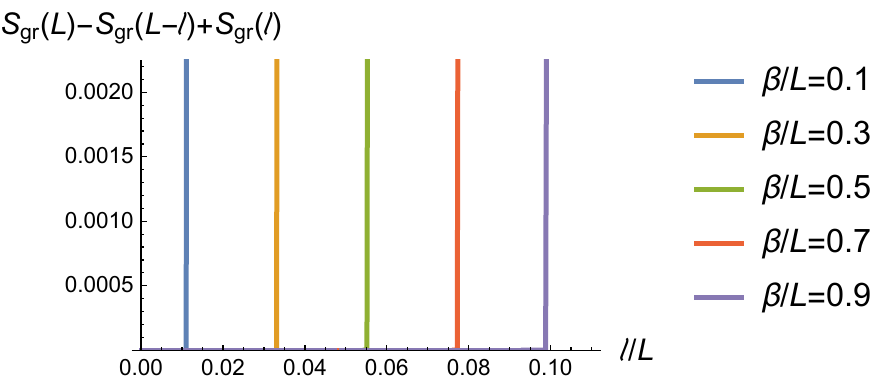}\label{hee2}}\\
  \caption{Holographic entanglement plateau in AdS$_3$/CFT$_2$. The Araki-Lieb inequality is saturated for a small enough critical length $\ell\leq\ell_c^\gr$. The figures are plotted in unit $c/3$. The right figure is the one being zoomed in around zero. It is obvious that there exists a critical length $\ell_c^\gr$, when $\ell$ is smaller than which $\D S$ vanishes. }\label{ee112}
\end{figure}

We consider a two-dimensional CFT on a circle of length $L$ and in a thermal state with inverse temperature $\b$. In this section we consider the high temperature case with  $\b\ll L$. We are interested in the single interval entanglement entropy.
From the Araki-Lieb inequality \cite{Araki:1970ba}, we know that
\be
|S(L-\ell)-S(\ell)| \leq S(L),
\ee
with $S(\ell)$, $S(L-\ell)$ being the entanglement entropies of the intervals with lengthes $\ell$ and $ (L-\ell)$ respectively and $S(L)$ being the thermal entropy of the system.
Holographically, it was found that  there exists a critical length $\ell_c^\gr$ so that when $\ell\leq \ell_c^\gr$, or equivalently when $\ell\geq L-\ell_c^\gr$, the Araki-Lieb inequality is saturated. The saturation is called the holographic entanglement plateau \cite{Hubeny:2013gta}. 
Indeed, the holographic entanglement entropy in this case is given by\cite{Azeyanagi:2007bj,Hubeny:2013gta}
\be \label{hh}
S_\gr(\ell) = \lt\{ \ba{ll}
\df{c}{3} \log \Big( \df{\b}{\pi\e} \sinh \df{\pi\ell}{\b} \Big)                        & ~\when~ \ell<L-\ell_c^\gr, \\
\df{c}{3} \log \Big( \df{\b}{\pi\e} \sinh \df{\pi(L-\ell)}{\b} \Big) + \df{\pi c L}{3\b} & ~\when~ \ell>L-\ell_c^\gr.
\ea \rt.
\ee
with
\be \label{lcgr}
\ell_c^\gr = \f{\b}{2\pi} \log \f{2}{1+\ep^{-2\pi L/\b}} \approx  \f{\b}{2\pi} \log 2.
\ee
The thermal entropy is holographically given by the Bekenstein-Hawking entropy of a non-rotating BTZ black hole
\be \label{SgrL}
S_\gr(L)=\f{\pi c L}{3\b}.
\ee
Then one can get the holographic entanglement plateau
\be
S_\gr(L) - S_\gr(L-\ell) + S_\gr(\ell) = 0, ~ \textrm{for} ~ \ell \leq \ell_c^\gr.
\ee
This has been shown in Figure~\ref{ee112}.

One implication of the entanglement plateau is that\cite{Azeyanagi:2007bj}
\be \label{cw}
\lim_{\ell \to 0} S(L-\ell) -S(\ell) = S(L).
\ee
This looks weaker than the holographic entanglement plateau, but it actually has interesting implications in 2D CFT. First of all, it has been proved to be true for any 2D CFT with a discrete spectrum \cite{Chen:2014ehg}. Secondly it makes sense at any temperature, not just the high temperature limit. However, the relation (\ref{cw}) generically holds only at strict $\ell \to 0$ limit. This limit may cover up many interesting points of the large interval entanglement entropy. Therefore in this work, we do not take this limit rigorously and focus on the quantity
\be
\D S=S(L) - S(L-\ell)+S(\ell).
\ee

\begin{figure}[htbp]
  \centering
  \subfigure[]{\includegraphics[height=40mm]{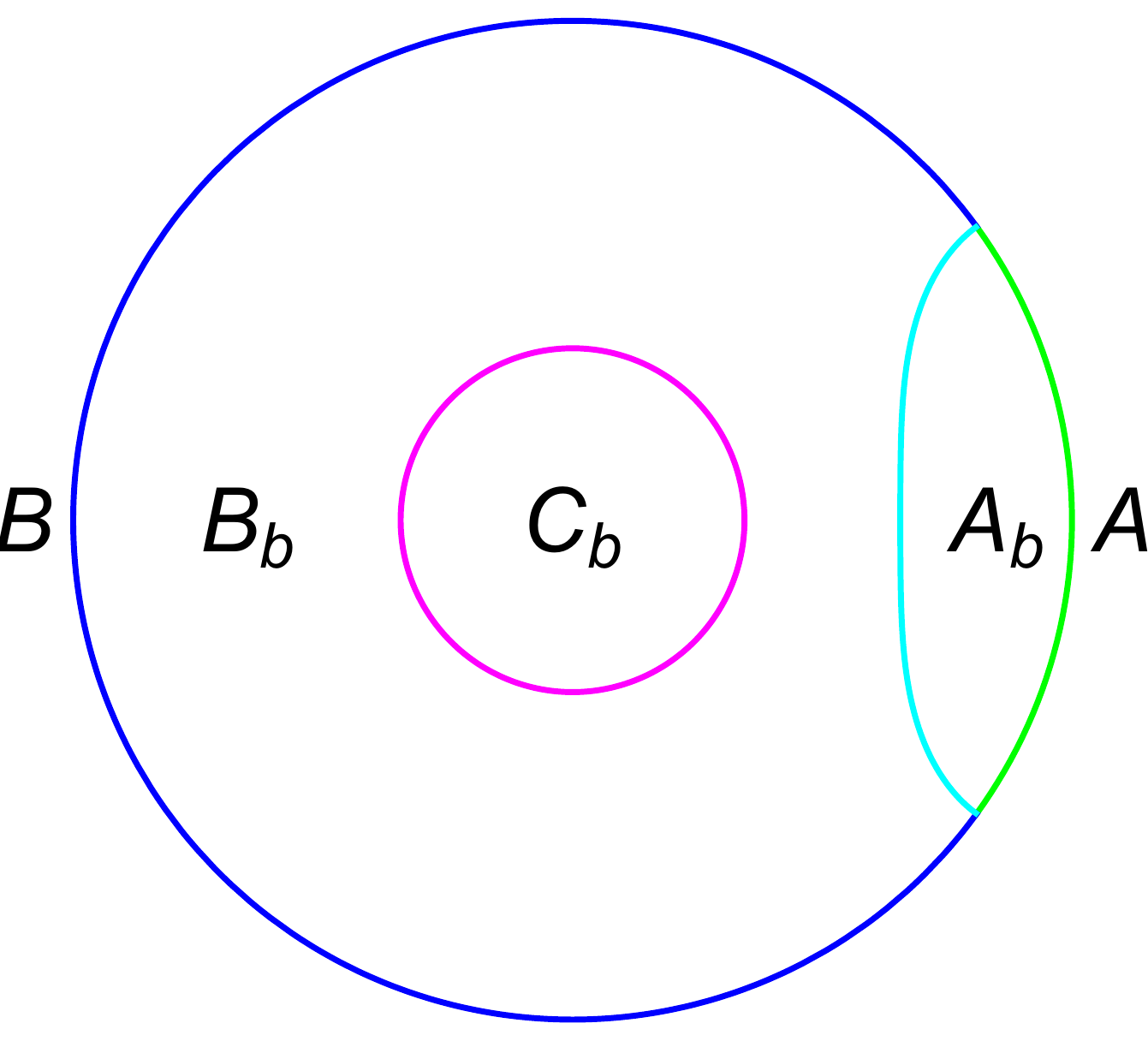}} ~~~~~~
  \subfigure[]{\includegraphics[height=40mm]{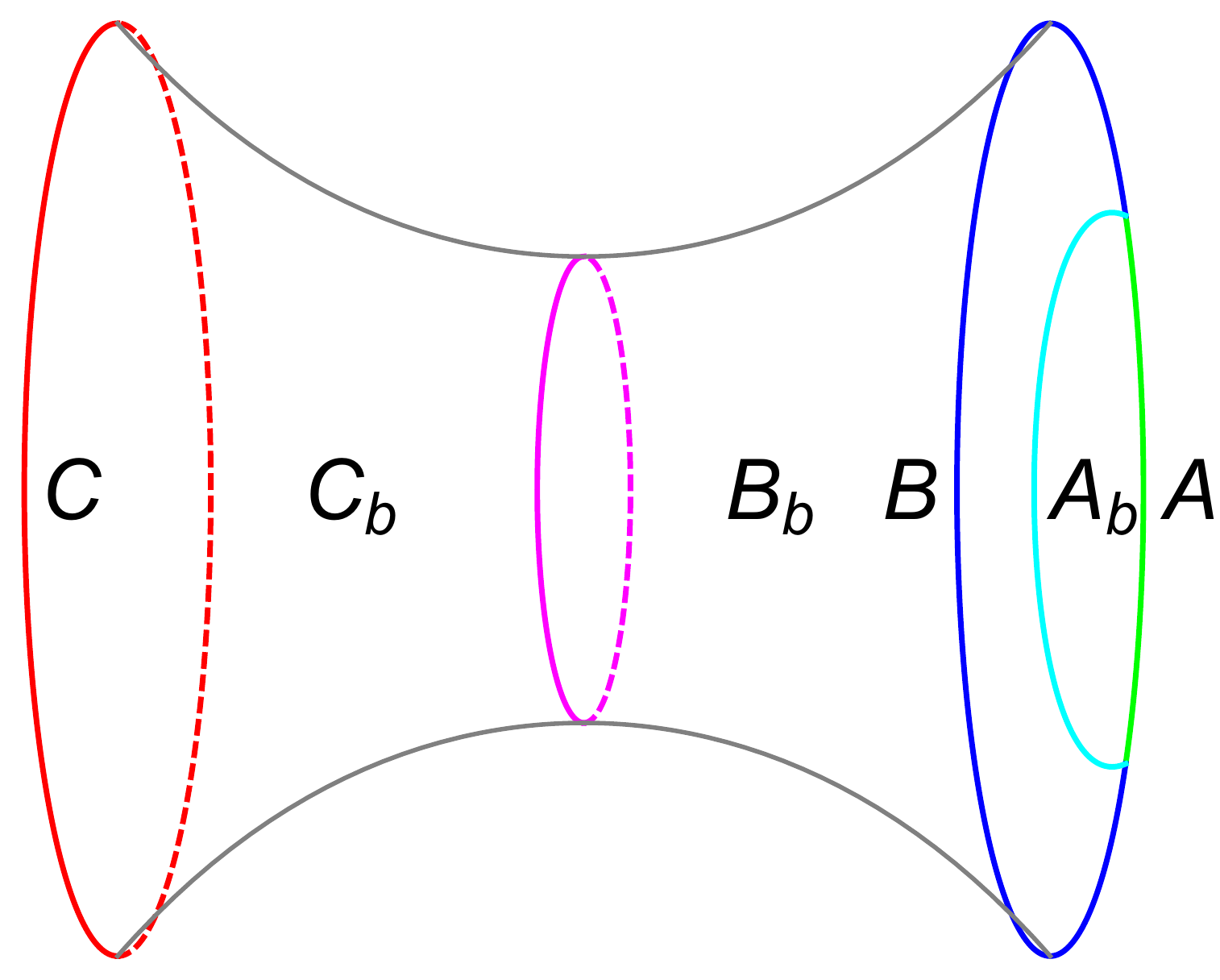}\label{hee2}}\\
  \caption{Corrections to the holographic entanglement plateau. They are captured by the mutual information $I(A_b,C_b)$ of two disconnected bulk regions \cite{Faulkner:2013ana}. Using the AdS/CFT correspondence, they also equal the mutual information $I(A,C)$ in CFT after the system is purified by adding the region $C$.}\label{hee12}
\end{figure}

On the gravity side, the holographic entanglement entropy is just the leading order classical contribution. The quantum correction to the holographic entanglement entropy has been discussed in \cite{Fujita:2009kw,Headrick:2010zt,Barrella:2013wja,Faulkner:2013ana}.  Especially in \cite{Faulkner:2013ana} by identifying the holographic entanglement entropy as the bulk entanglement entropy, one can get \cite{Faulkner:2013ana}
\be \label{flm}
S_{\gr}(L) - S_{\gr}(L-\ell)+S_{\gr}(\ell) = I(A_b,C_b) > 0.
\ee
As shown in Figure~\ref{hee12}, $I(A_b,C_b)$ is the mutual information between the bulk region $A_b$ and the black hole interior $C_b$, which is strictly positive as long as the size of $\ell$ is not identically zero.

The bulk eternal black hole is dual to the thermo-field double state, which could be taken as the purification of the thermal state.
As shown in Figure~\ref{hee2}, the whole system $A\cup B$ is in a thermal state, and the addition of another region $C$ makes the new system $A \cup B \cup C$ be in a pure state. Then we get
\be \label{cz}
\D S=S(L) - S(L-\ell)+S(\ell) = I(A,C) > 0,
\ee
with $I(A,C)$ being the mutual information between $A$ and $C$. Holographically, the mutual information between $A$ and $C$ is given by the mutual information between $A_b$ and $C_b$. Obviously, if one takes into account of the quantum correction, the Araki-Lieb inequality cannot be saturated \cite{Faulkner:2013ana}.

Next we would like to compute the quantity $\D S$ in the large $c$ CFT, which tells us the mutual  information between two bulk regions. The difficult part is on the computation of the entanglement entropy of a large interval. In the next subsection, we show how to relate the problem with the double-interval entanglement entropy on the complex plane, after omitting the exponentially suppressed terms proportional to the powers of $\ep^{-2\pi L/\b}$. This simplifies the discussion significantly.

\subsection{Long interval entanglement entropy}\label{sec2.2}

The R\'enyi entropy of a long interval with length $L-\ell$ in a CFT on a torus with spatial period $L$ and temporal period $\b$ has been discussed in \cite{Chen:2014ehg,Chen:2014hta,Chen:2015kua}.  The treatment therein applies to the case $\ell \ll \b \ll L$.
In this section we revisit the problem, and consider the case $\b \ll L$ and $\ell \ll L$ but we do not require $\ell \ll \b$. We omit the finite size corrections, which are the powers of $\ep^{-2\pi L/\b}$ so are exponentially suppressed.
More precisely, as discussed carefully in \cite{Chen:2015kua} such finite size corrections do not appear in the leading order entanglement entropy in the large $c$ limit but do appear in the R\'enyi entropies.
This allows us to consider only the contribution of the vacuum in the twist sector.
In other words, we may just consider the single interval entanglement entropy on a cylinder with period $\b$.
We show that the mutual information in (\ref{flm}), or equivalently in (\ref{cz}), equals the mutual information of two intervals on the complex plane.

\begin{figure}[htbp]
  \centering
  \includegraphics[width=0.9\textwidth]{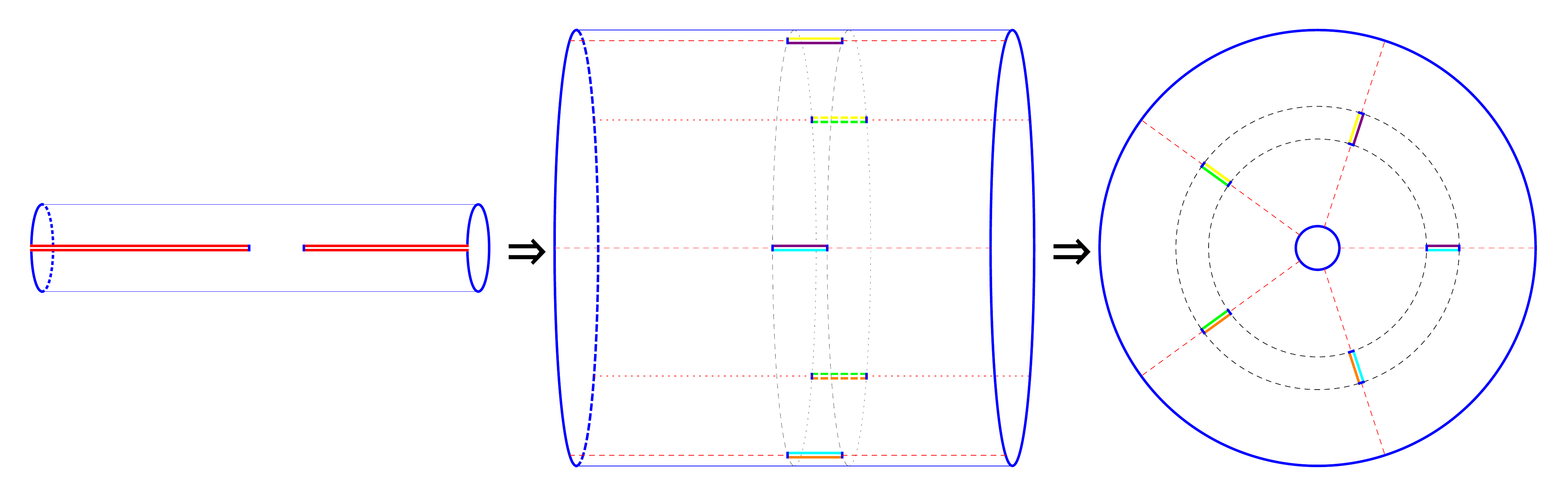}\\
  \caption{Illustration on the computation of long interval R\'enyi entropy. In the left figure we have the original cylinder $\mR$ with a length $L$ and a temporal period $\b$.  The long interval $[-L/2,v]\cup[u,L/2]$ with $u-v=\ell \ll L$ has length $L-\ell$.
   Via the replica trick and deforming the interval we get a cylinder $\mR_n$ with a length $L$, a temporal period $n\b$ and $n$ cuts, as shown in the middle figure. By a conformal transformation, the cylinder is transformed to an annulus, as shown in the right figure. In the last two figures, we take $n=5$.}\label{cyl}
\end{figure}

As shown in the left figure of Figure~\ref{cyl}, we consider the the long interval $A=[-L/2,v]\cup[u,L/2]$ with $\b \ll L$ and $u-v = \ell \ll L$. Via the replica trick we need to compute the partition function of the CFT on a  Riemann surface $\mR_n$, which is obtained by pasting $n$ torus along the cuts. In the limits $\b\ll L$, $\ell \ll L$, the torus is approximately a cylinder which we also denote by $\mR_n$, and for $n=1$ it is an ordinary cylinder $\mR$. As shown in the middle figure of Figure~\ref{cyl}, the cylinder $\mR_n$ now is of length $L$ and a temporal period $n\b$. We use the coordinate $w=x+\ii\t$ on $\mR_n$. There are $n$ cuts $[v + \ii j\b,u + \ii j\b]$, $j=0,1,\cdots,n-1$ with the same length $u-v=\ell$ in $\mR_n$, and the edges with the same color should be identified. This is due to the fact that one may deform the interval on the torus\cite{Chen:2014ehg}. The original interval is very large, almost along the whole spatial direction of the torus. We may take the interval to be the whole spatial direction minus the complement part, a short interval of length $\ell$. The presence of the interval along the spatial direction is not trivial. It induce the identification of the field in different replica such that the field theory is defined on a cylinder with a temporal period $n\b$ and $n$ short cuts of length $\ell$. The R\'enyi entropy is
\be \label{Sncyl}
S_n(L-\ell) \approx -\f{1}{n-1} \log \f{Z[\mR_n]}{Z[\mR]^n}.
\ee
In the above approximation, we have omitted the exponentially suppressed terms so that the partition functions are defined on the cylinder. The Riemann surface $\mR_n$ with coordinate $w$ can be mapped to an annulus with coordinate $z$ by the conformal transformation
\be
z=\ep^{\f{2\pi w}{n\b}}.
\ee
We denote the resulting annulus as $\mS_n$.
The $n$ cuts on $\mR_n$ are mapped to the cuts on the annulus along
\be \label{gj2}
[z_2^{(j)},z_1^{(j)}]=\big[\ep^{\f{2\pi v}{n\b}+\f{2\pi \ii j}{n}}, \ep^{\f{2\pi u}{n\b}+\f{2\pi \ii j}{n}}\big], ~ j=0,1,\cdots,n-1.
\ee
The boundaries of the cylinder at $x=\pm L/2$  are mapped to the boundaries of the annulus
\be \label{gj1}
|z|=\ep^{\pm\f{\pi L}{n\b}}.
\ee
In the right figure of Figure~\ref{cyl}, we show the annulus $\mS_n$ with $n$ cuts, the edges of the same color should be identified.
We have the partition function
\be
Z[\mR_n] = Z[\mS_n].
\ee
To regularize the ultra-violet(UV) divergences in the partition function and the R\'enyi entropy, we have to impose cutoffs at the boundaries of the cuts in $\mR_n$ and $\mS_n$. On $\mR_n$ we use the cutoff $\e$ for every  boundary, and so on $\mS_n$ we have the cutoffs
\be \label{gj3}
\f{2\pi\e}{n\b}z_1, ~ \f{2\pi\e}{n\b}z_2,
\ee
 for the boundaries $z_1^{(j)}$, $z_2^{(j)}$ respectively.

\begin{figure}[htbp]
  \centering
  \includegraphics[width=0.9\textwidth]{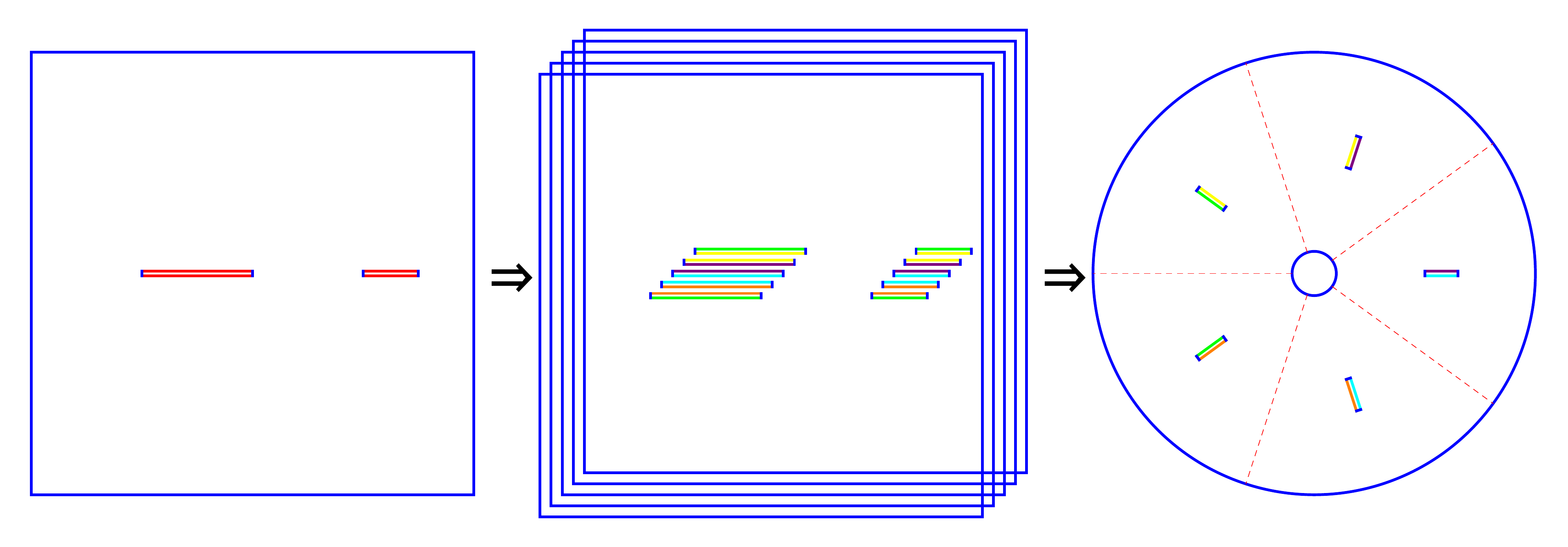}\\
  \caption{Illustration on the computation of the R\'enyi entropy of double intervals on the  complex plane. In the left figure we have the original complex plane $\mC$ with the coordinates $w$ and cuts $[w_4,w_3]\cup[w_2,w_1]$.
   Via the replica trick we get the $n$-fold complex plane $\mC_n$ in the middle figure.
  For the $n$-fold complex plane with the coordinate $w$, we may map it into an annulus $\td\mS_n$ with $n$ cuts, as shown in the right figure. Here we take $n=5$ as well.
    }\label{pl}
\end{figure}

Though we do not know how to calculate the partition function $Z[\mS_n]$ directly, we now show that it is related to the  R\'enyi entropy of double intervals on the complex plane.  We consider two intervals $[w_4,w_3]\cup[w_2,w_1]$ on the complex plane $\mC$ with the cross ratio
\be
x = \f{z_{12}z_{34}}{z_{13}z_{24}},
\ee
where $z_{ij} \equiv z_i-z_j$.
Via the replica trick we get the $n$-fold complex plane $\mC_n$, and the R\'enyi entropy
\be
S_n^\pl = -\f{1}{n-1} \log \f{Z[\mC_n]}{Z[\mC]^n}.
\ee
For the $\mC_n$ with the coordinate $w$, we can get the Riemann surface $\td\mS_n$ with the coordinate $z$ by using the transformation
\be
z= \Big( \f{w-w_3}{w-w_4} \Big)^{\f1{n}}.
\ee
Then we have the identification of the partition function
\be \label{Snpl}
Z[\mC_n] = Z[\td\mS_n].
\ee
For the boundary cuts $w_{1,2,3,4}^{(j)}$, $j=0,1,\cdots,n-1$ in $\mC_n$ we use different cutoffs $\e_{1,2,3,4}$ respectively. We find that $\td\mS_n$ turns out to be an annulus with boundaries at
\be \label{jj1}
|z|=\Big(\f{w_{34}}{\e_4}\Big)^{1/n}, ~ |z|=\Big(\f{\e_3}{w_{34}}\Big)^{1/n}.
\ee
There are $n$ cuts on the annulus, locating along
\be \label{jj2}
[z_2^{(j)},z_1^{(j)}]=\Big[\Big(\f{w_{23}}{w_{24}}\Big)^{1/n}\ep^{\f{2\pi\ii j}{n}},\Big(\f{w_{13}}{w_{14}}\Big)^{1/n}\ep^{\f{2\pi\ii j}{n}}\Big], ~~ j=0,1,\cdots,n-1,
\ee
and at $z_1^{(j)}$, $z_2^{(j)}$ there are cutoffs
\be \label{jj3}
\f{w_{34}\e_1}{n w_{13}w_{14}}z_1, ~ \f{w_{34}\e_2}{n w_{23}w_{24}}z_2.
\ee
This is shown in  Figure~\ref{pl}.

The annulus $\td\mS_n$ is in fact the same as $\mS_n$ with different parametrizations. After identifying  the boundaries (\ref{gj1})  (\ref{jj1}),  the cuts (\ref{gj2}), (\ref{jj2}), as well as the cutoffs at the cut boundaries (\ref{gj3}), (\ref{jj3}), we get the relations
\bea \label{id}
&& \e_1 = \f{2\pi\e}{\b} \f{w_{13}w_{14}}{w_{34}}, ~~
   \e_2 = \f{2\pi\e}{\b} \f{w_{23}w_{24}}{w_{34}}, \nn\\
&& \e_3 = \e_4 = w_{34} \ep^{-\f{\pi L}{\b}}, ~~
   \f{w_{13}}{w_{14}} = \ep^{\f{2\pi u}{\b}}, ~~
   \f{w_{23}}{w_{24}} = \ep^{\f{2\pi v}{\b}}.
\eea
Therefore we show that the R\'enyi entropy of a long interval on a torus with the high temperature (\ref{Sncyl}) equals approximately the R\'enyi entropy of two intervals on the  complex plane
\be
S_n(L-\ell) \approx S_n^\pl.
\ee
The approximation is exact up to the finite size correction, which is exponentially suppressed by the powers of $\ep^{-2\pi L/\b}$.
Note that for the entanglement entropy the approximation is actually exact for the leading order in the large $c$ limit, as there is no finite size correction in the $n\to 1$ limit \cite{Chen:2015kua}.


The R\'enyi entropy of two intervals on the complex plane (\ref{Snpl})  can be calculated by the correlation function of the twist operators \cite{Calabrese:2009ez,Headrick:2010zt,Calabrese:2010he,Chen:2013kpa,Chen:2013dxa}
\be
S_n^\pl = \f{c(n+1)}{6n} \log \f{w_{12}w_{34}}{(\e_1\e_2\e_3\e_4)^{1/2}} - I_n \Big(\f{z_{12}z_{34}}{z_{13}z_{24}}\Big).
\ee
Here $I_n (x)$ is the R\'enyi mutual information between the two intervals, and one can take $n\to1$ limit to read the mutual information $I(x)$.
In general, the mutual information $I(x)$ depends on the spectrum and the structure constants of the CFT.  For a large $c$ CFT, the contributions are dominated by the those from the vacuum module. We review the results in Appendix~\ref{appA}.
Using the identifications (\ref{id}), we get
\be \label{key}
S_n(L-\ell) \approx  \f{c(n+1)}{6n} \log \Big( \f{\b}{\pi\e} \sinh\f{\pi\ell}{\b} \Big)
                   + \f{\pi c(n+1)}{6n} \f{L}{\b}
                   - I_n(1-\ep^{-\f{2\pi\ell}{\b}}).
\ee
Note that we need $\b,\ell \ll L$ for the above approximation to be valid.
When $\ell \ll \b \ll L$, it is just
\be
S_n(L-\ell) \approx  \f{c(n+1)}{6n} \log \Big( \f{\b}{\pi\e} \sinh\f{\pi\ell}{\b} \Big)
                   + \f{\pi c(n+1)}{6n} \f{L}{\b}.
\ee
and this is in accord to the holographic entanglement entropy (\ref{hh}) and the results in \cite{Chen:2014ehg,Chen:2014hta,Chen:2015kua}.
When $\b \ll \ell \ll L$, the mutual information $I_n$ in (\ref{key}) gives an order $c$ contribution, which should be taken into account into the leading order contribution. Finally we get
\be
S_n(L-\ell) \approx  \f{c(n+1)}{6n} \log \Big( \f{\b}{\pi\e} \sinh\f{\pi(L-\ell)}{\b} \Big),
\ee
 which is in accord to the holographic entanglement entropy (\ref{hh}).

 It is remarkable that the treatment in this section has a larger validity domain than that in \cite{Chen:2014ehg,Chen:2014hta,Chen:2015kua}. An important simplification in our discussion is to omit the finite size correction, which include the exponentially suppressed terms. This allows us to get the result in the region $\b\ll \ell \ll L$, which is beyond the one in the existing treatment.

Another remarkable fact is that the terms proportional to $c$ in the entanglement entropies are actually of universal form. They are either the single-interval R\'enyi entropy at a finite temperature, or the R\'enyi entropy of the whole system. They are still true even for a general 2D CFT.

\subsection{Corrections to entanglement plateau}\label{sec2.4}
In the high temperature limit $\b \ll L$, the entanglement entropy of a short interval with $\ell/L \ll 1$ is approximately
\be \label{Ssh}
S_\sh(\ell) \approx \f{c}{3} \log \Big( \f{\b}{\pi\e} \sinh\f{\pi\ell}{\b} \Big).
\ee
Taking $n\to1$ limit of the result in the previous subsection we get the entanglement entropy of a long interval
\be
S_\lo(L-\ell) \approx  \f{c}{3} \log \Big( \f{\b}{\pi\e} \sinh\f{\pi \ell}{\b} \Big)
                   + \f{\pi c L}{3\b}
                   - I(1-\ep^{-\f{2\pi\ell}{\b}}).
\ee
For the large $c$ CFT, if we only consider the leading contributions, then using (\ref{miL}) we get
\be \label{Slo}
S_\lo(\ell) = \lt\{ \ba{ll}
\df{c}{3} \log \df{\b}{2\pi\e} + \df{c\pi\ell}{3\b}                    & ~\when~ \ell<L-\ell_c^\cft \\
\df{c}{3} \log \Big( \df{\b}{\pi\e} \sinh \df{\pi(L-\ell)}{\b} \Big) + \df{\pi c L}{3\b} & ~\when~ \ell>L-\ell_c^\cft,
\ea \rt.
\ee
with the critical length
\be
\ell_c^\cft = \f{\b}{2\pi}\log 2
\ee
which is the same as the gravity critical length $\ell_c^\gr$ (\ref{lcgr}) in high temperature limit.

\begin{figure}[htbp]
  \centering
  \subfigure[$\b/L=0.1$]{\includegraphics[height=27mm]{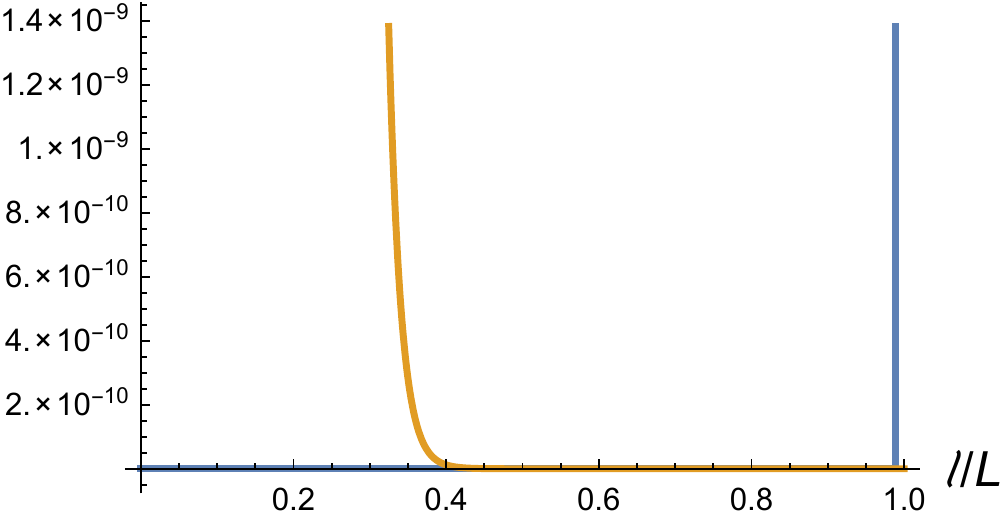}}
  \subfigure[$\b/L=0.3$]{\includegraphics[height=27mm]{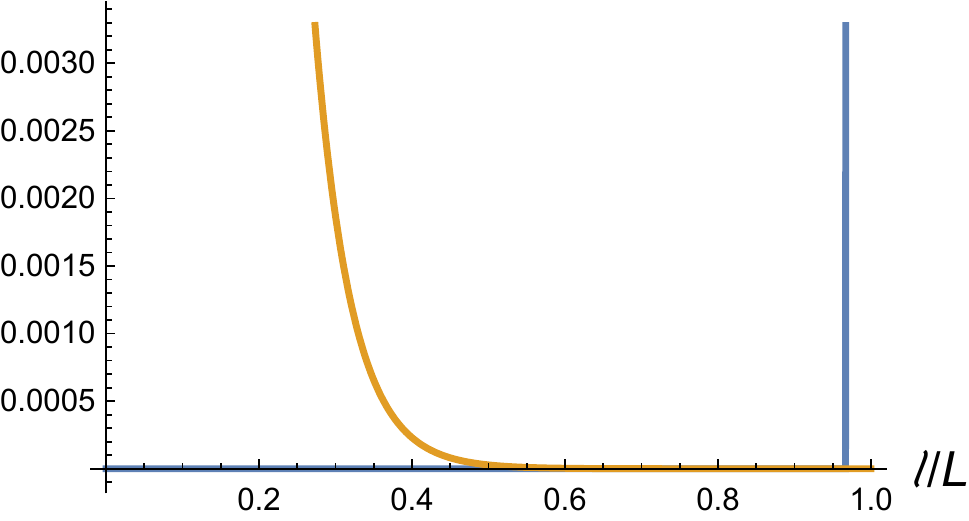}}
  \subfigure[$\b/L=0.5$]{\includegraphics[height=27mm]{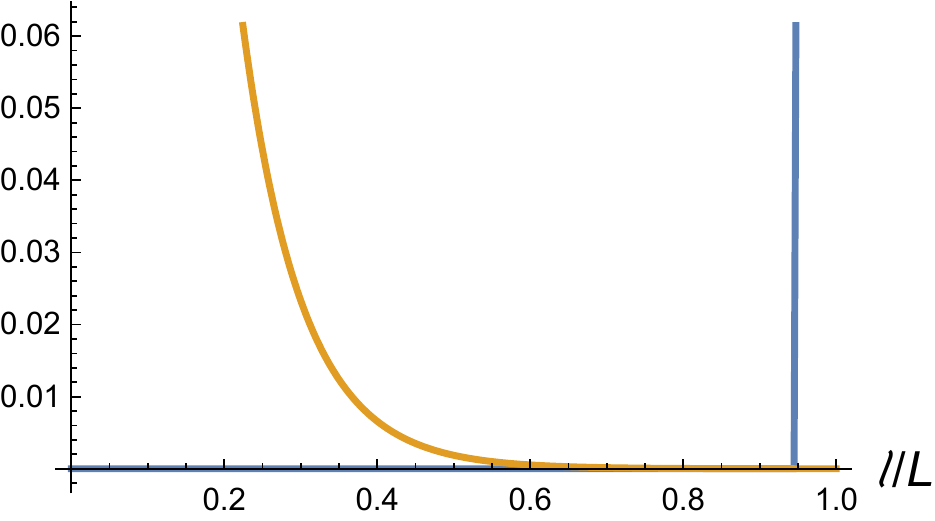}}\\
  \subfigure[$\b/L=0.7$]{\includegraphics[height=27mm]{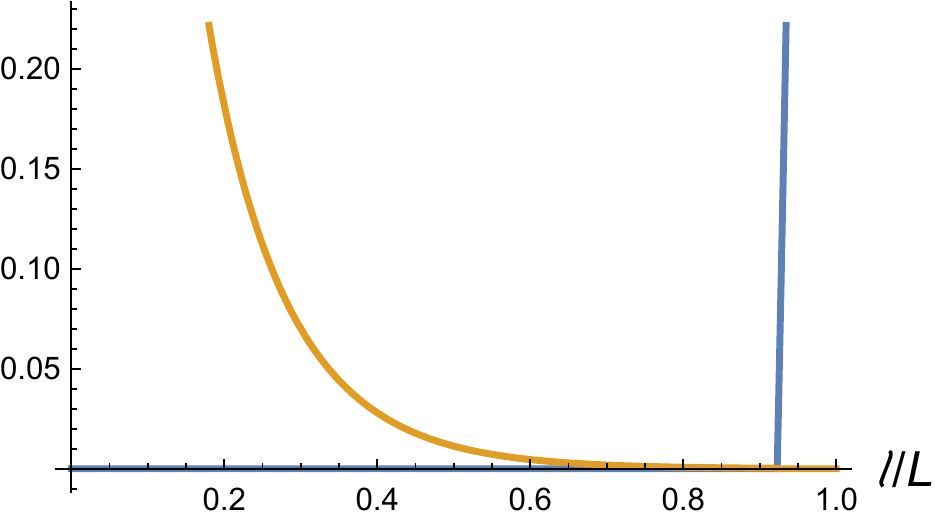}}
  \subfigure[$\b/L=0.9$]{\includegraphics[height=27mm]{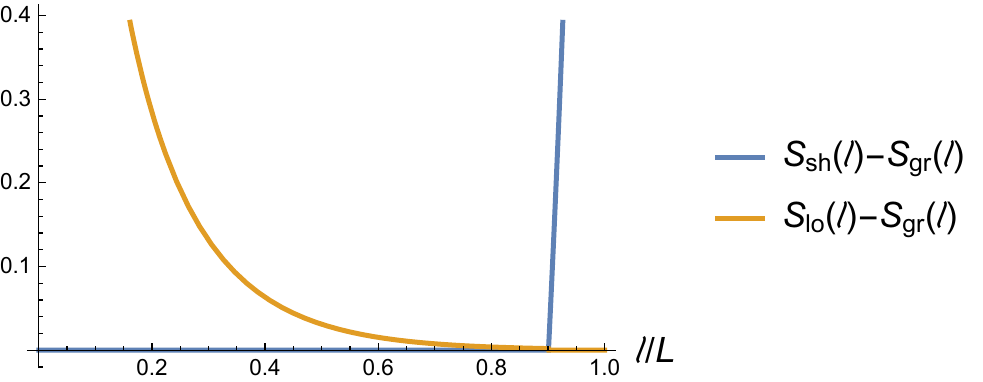}}
  \caption{In the large $c$ limit, the leading order entanglement entropy of a short interval (\ref{Ssh}) and a long interval (\ref{Slo}). We use the holographic entanglement entropy (\ref{hh}) as the benchmark to compare. The figures are plotted in unit of $c/3$.}\label{dS}
\end{figure}

Although the short interval entanglement entropy (\ref{Ssh}) is derived with the assumption $\ell/L \ll 1$, it has a much larger validity domain and matches the gravity result (\ref{hh}) as long as $0<\ell<L-\ell_c$. The long interval entanglement entropy (\ref{Slo}) is derived with assumption $(L-\ell)/L \ll 1$, and it strictly matches the gravity result (\ref{hh}) for $L-\ell_c < \ell <L$, and it also approximately matches (\ref{hh}) for $\b \ll \ell <L$. One can see this in Figure~\ref{dS}. Note that the short interval result (\ref{Ssh}) breaks down abruptly as $\ell\to L$, and and long interval result (\ref{Slo}) breaks down in a milder way as $\ell\to 0$.

\begin{figure}[htbp]
  \centering
  \subfigure[]{\includegraphics[height=35mm]{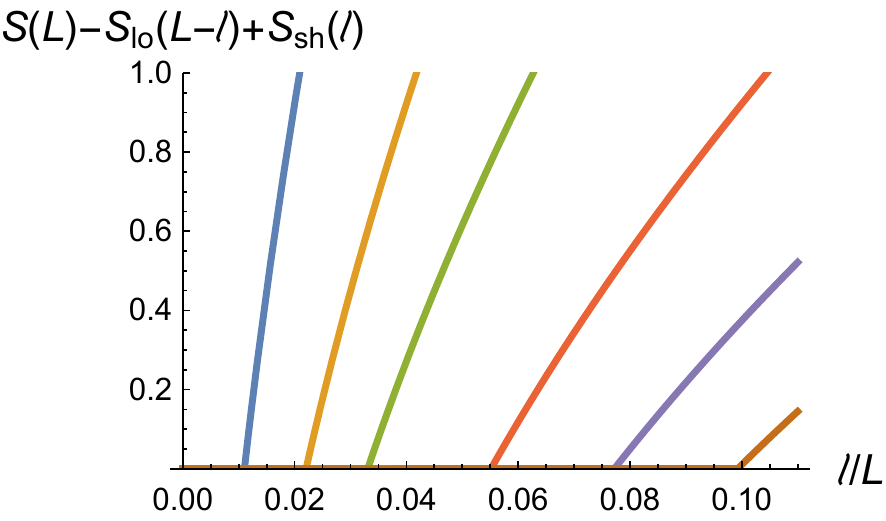}} ~~~
  \subfigure[]{\includegraphics[height=35mm]{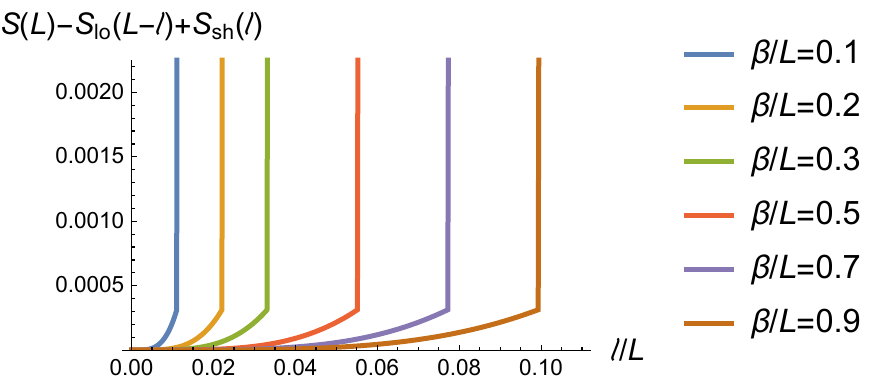}}
  \caption{Corrections to the entanglement plateau in 2D CFT. We just set $c=3$ to plot the figures. From the left figure it seems suggest that the plateau is still there. The right figure shows $\D S$ after zooming in around the zero. From the right figure, it is easy to see that the plateau disappears: $\D S=0$ only when $l \to 0$. }\label{ee212}
\end{figure}

Next we consider the next-to-leading order contribution to the entanglement entropies of the long and the short intervals in the large $c$ limit. We will see how such correction change the entanglement plateau.
First of all, for a CFT at a high temperature, omitting the exponentially suppressed terms, one can easily get its thermal entropy
\be
S(L) \approx \f{\pi c L}{3\b},
\ee
which equals the black hole entropy (\ref{SgrL}). Then the correction to the entanglement plateau  is
\be
\D S=S(L) - S_\lo(L-\ell) + S_\sh(\ell) = I(1-\ep^{-\f{2\pi\ell}{\b}}) >0,
\ee
which is strictly positive as long as $\ell \neq 0$. With the contributions from only the vacuum module, we use (\ref{miL}), (\ref{z92}) and plot it in Figure~\ref{ee212}, and one can compare it with the gravity result in Figure~\ref{ee112}.

There are other contributions to the mutual information from nonvacuum modules. As we have shown, the function $I(x)$ is actually related to the mutual information between two intervals. The contribution from other modules can be read in a straightforward way. In particular, as shown in \cite{Calabrese:2010he,Chen:2017hbk}, the leading contribution from a primary module could be of a universal form. As a result, when $\ell \ll \b$, the correction from a nonvacuum module with a primary operator $\mX$ of scaling dimension $\D_\mX$ takes a universal form so that we have
\be
 \d_\mX \big( S(L) - S_\lo(L-\ell) + S_\sh(\ell) \big) = \frac{\sqrt{\pi} \Gamma(2 \Delta_\mX +1) }{4\Gamma(2 \Delta_\mX  +3/2)} \Big(\f{\pi\ell}{\b}\Big)^{2\D_\mX}+ O(\ell^{2\D_\mX+1},\ell^{3\D_\mX}).
\ee
Note that the universal contribution from the nonvacuum module is independent of the central charge and the structure constants.
For the contributions from the nonvacuum modules, only the leading one from each module takes a universal form, while the subleading ones rely on the details of the theory.

\section{Low temperature case} \label{sec3}

In this section we consider the low temperature case. To make the equations concise, we only include the contributions of the holomorphic sector, and those from the anti-holomorphic sector can be added easily.

At a low temperature, the dual gravity configuration is the thermal AdS, and the holographic entanglement entropy is always
\be
S_\gr(\ell) = \f{c}{6}\log\Big(\f{L}{\pi\e}\sin\f{\pi\ell}{L}\Big).
\ee
One can see that $S_\gr(L-\ell)=S_\gr(\ell)$, and this leads to
\be
S_\gr(L-\ell)-S_\gr(\ell)=0.
\ee
This is consistent with the fact that the classical entropy of thermal AdS is vanishing
\be
S_\gr(L)=0.
\ee
The holographic entanglement plateau is trivial for the low temperature case
\be \label{pla2}
\D S=S_\gr(L) - | S_\gr(L-\ell) - S_\gr(\ell) | =0.
\ee

\begin{figure}[htbp]
  \centering
  \subfigure[]{\includegraphics[height=40mm]{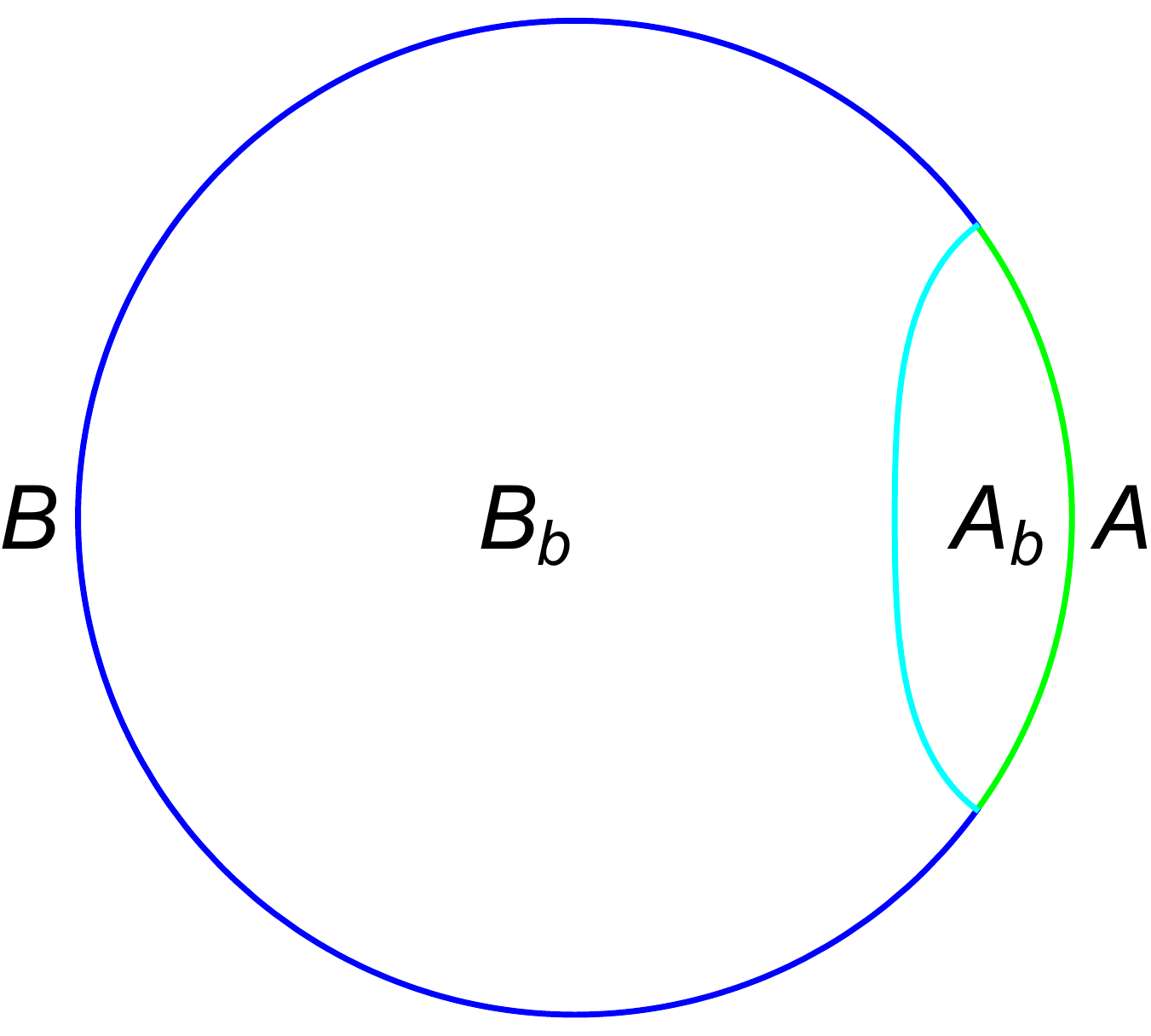}} ~~~~~~
  \subfigure[]{\includegraphics[height=40mm]{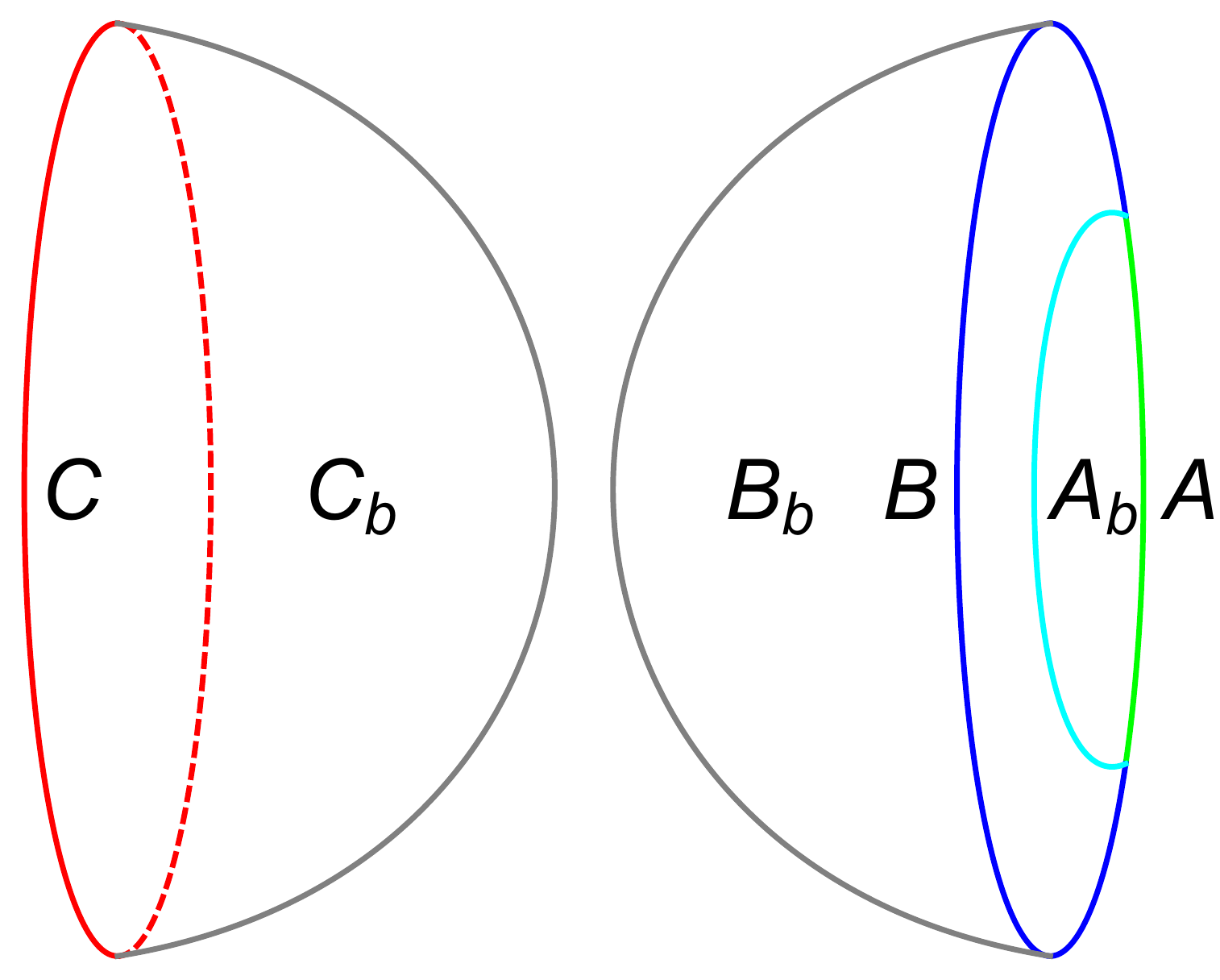}}\\
  \caption{Purification of the thermal density matrix in gravity. In (b) there is no macroscopic horizon that connects the two asymptotic AdS regions $A_b\cup B_c$ and $C_b$. However, there are still microscopic connections between $A_b\cup B_c$ and $C_b$, and the two boundaries $A\cup B$ and $C$. We have the mutual information $I(A_b,C_b)>0$ and $I(A,C)>0$.}\label{hee34}
\end{figure}

Although there is no horizon in the thermal AdS, the idea in \cite{Faulkner:2013ana} still applies. At the high temperature, the purification of the thermal density matrix leads to the thermo-field double state. Holographically there is the eternal black hole, in which the wormhole connecting two asymptotically AdS regions. At the low temperature, we do not have the eternal black hole picture, but we can still have the picture on purification of the thermal density matrix, see Figure~\ref{hee34}. Therefore, we still have the quantum corrections (\ref{flm}) and (\ref{cz}):
  \be
  \D S=S_\gr(L) - | S_\gr(L-\ell) - S_\gr(\ell) | =I(A_b,C_b)=I(A,C)>0.
  \ee
  Note that even at the low temperature, the thermal entropy is not strictly vanishing.
  In the following, we would like to compute $\D S$ to the next-to-leading order. The computation relies on the expansion of the thermal density matrix. In Appendix \ref{appC}, as a double check we use the OPE of the twist operators to compute the entanglement entropies in this section, and find good agreement. 

\subsection{Short interval entanglement entropy}\label{sec3.2}

Let us focus on the vacuum module, which is dominant in the large $c$ limit.
We revisit the contributions of the holomorphic stress tensor $T$ to the short interval entanglement entropy. In \cite{Cardy:2014jwa}, it was shown that the entanglement entropy is
\be \label{z1}
S(\ell) = \f{c}{6}\log\Big(\f{L}{\pi\e}\sin\f{\pi\ell}{L}\Big)
         + 4 q^2 \Big( 1 - \f{\pi\ell}{L} \cot \f{\pi\ell}{L} \Big)
         + O(q^3),
\ee
with
\be
q \equiv \ep^{-2\pi\b/L}.
 \ee
 It was believed that it applies to an interval of length $\ell$ as long as $\ell$ is not comparable to the size of the circle $L$. In fact the result is divergent in the limit $\ell \to L$. In this subsection, we give a more scrutinized derivation of the short interval entanglement entropy, and find a result that is consistent with (\ref{z1}). 

The un-normalized thermal density matrix could be expanded as
\be
\r = |0\rag\lag0| + \f{q^2}{\a_T}  |T\rag\lag T| + O(q^3),
\ee
with $\a_T=\f{c}{2}$, and so the reduced density matrix is
\be \label{z15}
\r_A = \r_{A,0} + \f{q^2}{\a_T} \r_{A,T} + O(q^3),
\ee
with
\be
\r_{A,0}=\tr_B |0\rag \lag0|, \hs{3ex}\r_{A,T}=\tr_B |T\rag \lag T|. \ee
The R\'enyi entropy is
\be
S_n(\ell) = -\f1{n-1} \log \f{\tr_A\r_A^n}{(\tr_A\r_A)^n},
\ee
with
\be
\tr_A\r_A = \tr\r = 1 + q^2 + O(q^3).
\ee
We organize $\tr_A\r_A^n$ by the expansion of $q^2$ as
\be \label{deff}
\f{\tr_A\r_A^n}{\tr_A\r_{A,0}^n} = f(n) + O(q^3)  = \sum_{k=0}^{n} q^{2k} f(n,k) + O(q^3).
\ee
Note that as we will take $n \to 1$ limit in the last,  now we keep all the terms of orders $q^2$, $q^4$, $\cdots$, $q^{2(n-1)}$, $q^{2n}$. This is different from the treatment in \cite{Cardy:2014jwa}.
In the following we use the subscript ``$_\sh$" to denote  the results for a short interval of length $\ell$, and the subscript ``$_\lo$" to denote the ones for  a long interval of length $L-\ell$.

It is known that the contribution from the vacuum is\cite{Calabrese:2004eu}
\be
\tr_A\r_{A,0}^n= \Big( \f{L}{\pi\e} \sin \f{\pi\ell}{L} \Big)^{-2h_\s},
\ee
with $h_\s$ being the conformal dimension of the twist operators
\be
h_\s= \f{c(n^2-1)}{24n}.
\ee
To compute the contributions $f_\sh(n,k)$ from the excitations in the vacuum module, we
 may  map the $n$-fold cylinder to a complex plane \cite{Cardy:2014jwa,Chen:2014unl}, and find
\bea \label{fnk}
&& f_\sh(n,k) = \sum_{0\leq j_1< \cdots<j_k\leq n-1}
            \f{(\f{2\ii}{n}\sin\f{\pi\ell}{L})^{4k}}{\a_T^k} \Big \lag
\prod_{a=1}^k \Big[
                \ep^{\f{4\pi\ii}{n}(2j_a+\f{\ell}{L})}
                T(\ep^{\f{2\pi\ii}{n}(j_a+\f{\ell}{L})})
                T(\ep^{\f{2\pi\ii}{n}j_a})
              \Big]
\Big\rag_\mC \nn\\
&& \phantom{f_\sh(n,k) =}
   + \f{n(n^2-1)}{12} \sum_{1\leq j_1< \cdots<j_{k-1}\leq n-1} \f{(\f{2\ii}{n}\sin\f{\pi\ell}{L})^{4k}}{\a_T^{k-1}} \Big \lag
   \Big[  \ep^{\f{4\pi\ii\ell}{nL}}T(\ep^{\f{2\pi\ii\ell}{nL}}) + T(0)  \Big]\nn\\
&& \phantom{f_\sh(n,k) =}
\times\prod_{a=1}^{k-1} \Big[
                \ep^{\f{4\pi\ii}{n}(2j_a+\f{\ell}{L})}
                T(\ep^{\f{2\pi\ii}{n}(j_a+\f{\ell}{L})})
                T(\ep^{\f{2\pi\ii}{n}j_a})
              \Big]\Big\rag_\mC + O(n-1)^2,
\eea
among which
\be \label{e39}
f_\sh(n,0) = 1, ~~
f_\sh(n,1) = n \Big( \f{\sin\f{\pi\ell}{L}}{n\sin\f{\pi\ell}{n L}} \Big)^{4} + O(n-1)^2.
\ee
Note that to calculate the entanglement entropy, we do not need the $O(n-1)^2$ term.
We get the short interval R\'enyi entropy
\be \label{e31}
S^\sh_n(\ell) = \f{c(n+1)}{12n}\log \Big( \f{L}{\pi\e}\sin\f{\pi\ell}{L} \Big)
+ \f{1}{n-1} \log \f{ ( 1 + q^2 + O(q^3))^n}
                    {\sum_{k=0}^n q^{2k} f_\sh(n,k) + O(q^3) }.
\ee

To calculate the entanglement entropy for the case at hand, we need to take four limits, i.e., the low temperature limit $q=\ep^{-2\pi\b/L} \to 0$, the large central charge limit $1/c \to 0$, the short interval limit $\ell \to 0$, and the $n \to 1$ limit. There may be subtleties in the order of taking the limits. We have taken the low temperature first. Since we do not know how to calculate the the R\'enyi entropy (\ref{e31}) for general length $\ell$, we will take the limit $1/c \to 0$, and then $\ell \to 0$ before taking $n \to 1$. We just assume that the chosen order of taking the limits  does not affect the final result.


Noting that
\be
\lim_{\ell\to0}f_\sh(n,k)=C_n^k, ~~ \lim_{\ell\to0}f_\sh(n)= (1+q^2)^n,
\ee
we read
\be \label{lim1}
\lim_{\ell \to 0} S^\sh_n(\ell) = \f{c(n+1)}{12n}\log \f{\ell}{\e} + O(q^3).
\ee
We define
\be \label{e32}
a(n) \equiv (1+q^2)^n - f_\sh(n) = \sum_{k=0}^n q^{2k} a(n,k),
\ee
and further write it as
\be \label{z2}
a(n)= a_\I(n)+a_\II(n)+a_\III(n),~~a(n,k) = a_\I(n,k) + a_\II(n,k) + a_\III(n,k).
\ee
Explicitly, we have
\bea \label{z3}
&&\hspace{-9mm} a_\I(n,k) = C_n^k \Big( 1 - \Big( \f{\sin\f{\pi\ell}{L}}{n\sin\f{\pi\ell}{n L}} \Big)^{4k} \Big), \nn\\
&&\hspace{-9mm} a_\II(n,k) =     \sum_{0\leq j_1 < \cdots <j_{k} \leq n-1}
                    \Big( \f{\sin\f{\pi\ell}{L}}{n\sin\f{\pi\ell}{n L}} \Big)^{4k}
                    \Big\lag 1 -
                    \f{(2\ii\sin\f{\pi\ell}{n L})^{4k}}{\a_T^k}\prod_{a=1}^k
                             \Big[ \ep^{\f{4\pi\ii}{n}(2j_a+\f{\ell}{L})}
                                   T(\ep^{\f{2\pi\ii}{n}(j_a+\f{\ell}{L})})
                                   T(\ep^{\f{2\pi\ii}{n}j_a}) \Big]
                    \Big\rag_\mC, \nn\\
&&\hspace{-9mm} a_\III(n,k) = - \f{n(n^2-1)}{12}
     \sum_{1\leq j_1< \cdots<j_{k-1}\leq n-1} \Big\{
     \Big( \f{\sin\f{\pi\ell}{L}}{n\sin\f{\pi\ell}{n L}} \Big)^{4(k-1)}
     \Big(\f{2\ii}{n}\sin\f{\pi\ell}{L}\Big)^4
     \Big\lag
     \Big[  \ep^{\f{4\pi\ii\ell}{nL}}T(\ep^{\f{2\pi\ii\ell}{nL}}) + T(0)  \Big]\nn\\
&&\hspace{-9mm} \phantom{g_\III(n,k) =} \times
     \f{(2\ii\sin\f{\pi\ell}{n L})^{4(k-1)}}{\a_T^{k-1}}\prod_{a=1}^{k-1}
                             \Big[ \ep^{\f{4\pi\ii}{n}(2j_a+\f{\ell}{L})}
                                   T(\ep^{\f{2\pi\ii}{n}(j_a+\f{\ell}{L})})
                                   T(\ep^{\f{2\pi\ii}{n}j_a}) \Big]
     \Big\rag_\mC \Big\}.
\eea
Putting (\ref{e32}) in (\ref{e31}) and taking the limit $n\to1$, we get the short interval entanglement entropy
\be
S_\sh(\ell) = \f{c}{6} \log \Big( \f{L}{\pi\e}\sin\f{\pi\ell}{L} \Big) + \f{a'(1)+O(q^3)}{1+q^2+O(q^3)},
\ee
with
\be
a'(1) = \p_n a(n)|_{n=1} = a'_\I(1) + a'_\II(1) + a'_\III(1).
\ee
Using (\ref{z3}) we can easily get
\bea \label{e42}
&& a_\I(n) = (1+q^2)^n - \Big( 1 +q^2 \Big( \f{\sin\f{\pi\ell}{L}}{n\sin\f{\pi\ell}{n L}} \Big)^{4} \Big)^n, \nn\\
&& a'_\I(1) = 4 q^2 \Big( 1 - \f{\pi\ell}{L} \cot \f{\pi\ell}{L} \Big).
\eea

Note that $a_\II(n)$, $a_\III(n)$ defined in (\ref{z2}), (\ref{z3}) are at least of order $q^2$, we get the short interval entanglement entropy
\be \label{e38}
S_\sh(\ell) = \f{c}{6} \log \Big( \f{L}{\pi\e}\sin\f{\pi\ell}{L} \Big)
               + 4 q^2 \Big(  1 - \f{\pi\ell}{L} \cot \f{\pi\ell}{L} \Big)
               + a'_\II(1) + a'_\III(1) +O(q^3).
\ee
We cannot evaluate $a'_\II(1)$ or $a'_\III(1)$ explicitly for general $c$ or general $\ell$.  We may expand the entanglement entropy in powers of $1/c$ and $\ell$. Then the  order $c^0$ part of $a_\II(n)$ is of order $\ell^8$, and the order $c^0$ part of $a_\III(n)$ is of  order $\ell^6$. So we get that
\be \label{e67}
S_\sh(\ell) = \f{c}{6} \log \Big( \f{L}{\pi\e}\sin\f{\pi\ell}{L} \Big)
              + 4 q^2 \Big(  1 - \f{\pi\ell}{L} \cot \f{\pi\ell}{L} \Big)
              + O(q^3,1/c,\ell^6).
\ee
Here $O(q^3,1/c,\ell^6)$ is schematic. It can denote the terms that are of order $q^3$, no mater what orders the terms are in the expansion of $1/c$, $\ell$. It may also denote the terms that are of order $1/c$, no mater what orders they are in the expansion of $q$, $\ell$, or denote the terms that are of order $\ell^6$, no mater what orders in the expansion of $q$, $1/c$. This is consistent with (\ref{z1}), and is in fact a much relaxed version. 

\subsection{Long interval entanglement entropy}\label{sec3.3}

 The result (\ref{z1}) is divergent in $\ell \to L$ limit, and this suggests that for a long interval, the entanglement entropy should be reconsidered carefully. We keep $\ell$ to be small and take $L-\ell$ to be large. The above computation still make sense
 and we need to set $\ell \to L-\ell$ in (\ref{fnk}) to get the result for a  long interval
\bea \label{z6}
&& f_\lo(n,k) = \sum_{0\leq j_1< \cdots<j_k\leq n-1}
            \f{(\f{2\ii}{n}\sin\f{\pi\ell}{L})^{4k}}{\a_T^k} \Big \lag
\prod_{a=1}^k \Big[
                \ep^{\f{4\pi\ii}{n}(2j_a+1-\f{\ell}{L})}
                T(\ep^{\f{2\pi\ii}{n}(j_a+1-\f{\ell}{L})})
                T(\ep^{\f{2\pi\ii}{n}j_a})
              \Big]
\Big\rag_\mC \nn\\
&& \phantom{f(n,k) =}
   + \f{n(n^2-1)}{12} \sum_{1\leq j_1< \cdots<j_{k-1}\leq n-1} \f{(\f{2\ii}{n}\sin\f{\pi\ell}{L})^{4k}}{\a_T^{k-1}} \Big \lag
   \Big[  \ep^{\f{4\pi\ii}{n}(1-\f{\ell}{L})}T(\ep^{\f{2\pi\ii}{n}(1-\f{\ell}{L})}) + T(0)  \Big]\nn\\
&& \phantom{f(n,k) =}
\times\prod_{a=1}^{k-1} \Big[
                \ep^{\f{4\pi\ii}{n}(2j_a+1-\f{\ell}{L})}
                T(\ep^{\f{2\pi\ii}{n}(j_a+1-\f{\ell}{L})})
                T(\ep^{\f{2\pi\ii}{n}j_a})
              \Big]\Big\rag_\mC + O(n-1)^2.
\eea
As the results (\ref{e39}) still apply, after sending $\ell \to L-\ell$ we get
\be
f_\lo(n,0) = 1, ~~
f_\lo(n,1) = n \Big( \f{\sin\f{\pi\ell}{L}}{n\sin\f{\pi(L-\ell)}{n L}} \Big)^{4} + O(n-1)^2,
\ee
and $f_\lo(n,n)=f_\sh(n,n)$.
Therefore we get the long interval R\'enyi entropy
\be \label{e40}
S^\lo_n(L-\ell) = \f{c(n+1)}{12n}\log \Big( \f{L}{\pi\e}\sin\f{\pi\ell}{L} \Big)
+ \f{1}{n-1} \log \f{ ( 1 + q^2 + O(q^3))^n}
                    {\sum_{k=0}^n q^{2k} f_\lo(n,k) + O(q^3) }.
\ee

In the $\ell \to 0 $ limit, we have
\bea
&& \lim_{\ell\to0} f_\lo(n,k) =0 ~~~\for~ k = 1,2,\cdots,n-1, \nn\\
&& \lim_{\ell\to0} f_\lo(n,n) = 1, ~~ \lim_{\ell\to0} f_\lo(n) = 1 + q^{2n},
\eea
Let us define
\be \label{e41}
b(n) \equiv 1+q^{2n} - f_\lo(n) = b_\I(n) + b_\II(n) + b_\III(n) + b_\IV(n),
\ee
with
\bea \label{z7}
&& b_\I(n) = q^{2n} \Big( 1- \Big( \f{\sin\f{\pi\ell}{L}}{n\sin\f{\pi\ell}{n L}} \Big)^{4n} \Big),\nn\\
&& b_\II(n) = q^{2n} \Big( \f{\sin\f{\pi\ell}{L}}{n\sin\f{\pi\ell}{n L}} \Big)^{4n}
                  \Big\lag 1 -
                      \f{(2\ii\sin\f{\pi\ell}{n L})^{4n}}{\a_T^n}\prod_{j=0}^{n-1}
                             \Big[ \ep^{\f{4\pi\ii}{n}(2j+\f{\ell}{L})}
                                   T(\ep^{\f{2\pi\ii}{n}(j+\f{\ell}{L})})
                                   T(\ep^{\f{2\pi\ii}{n}j}) \Big]
                  \Big\rag_\mC, \nn\\
&& b_\III(n) =  - \f{n(n^2-1)}{12}q^{2n}
     \Big( \f{\sin\f{\pi\ell}{L}}{n\sin\f{\pi\ell}{n L}} \Big)^{4(n-1)}
     \Big(\f{2\ii}{n}\sin\f{\pi\ell}{L}\Big)^4
     \Big\lag
     \Big[  \ep^{\f{4\pi\ii\ell}{nL}}T(\ep^{\f{2\pi\ii\ell}{nL}}) + T(0)  \Big] \nn\\
&& \phantom{h_\III(n) =} \times
     \f{(2\ii\sin\f{\pi\ell}{n L})^{4(n-1)}}{\a_T^{n-1}}\prod_{j=1}^{n-1}
                             \Big[ \ep^{\f{4\pi\ii}{n}(2j+\f{\ell}{L})}
                                   T(\ep^{\f{2\pi\ii}{n}(j+\f{\ell}{L})})
                                   T(\ep^{\f{2\pi\ii}{n}j}) \Big]
     \Big\rag_\mC
   + O(n-1)^2, \nn\\
&& b_\IV(n) = -\sum_{k=1}^{n-1} q^{2k} f_\lo(n,k).
\eea
Putting (\ref{e41}) into (\ref{e40}) and taking $n \to 1$ limit, we get the long interval entanglement entropy
\be \label{e43}
S_\lo(L-\ell) = \f{c}{6} \log \Big( \f{L}{\pi\e}\sin\f{\pi\ell}{L} \Big) + S(L) + \f{b'(1)+O(q^3)}{1+q^2+O(q^3)},
\ee
with
\be
b'(1) = \p_n b(n)|_{n=1} = b'_\I(1) + b'_\II(1) + b'_\III(1) + b'_\IV(1).
\ee
Here we have the thermal entropy
\be\label{e76}
S(L) = \Big( 1 + \f{4\pi \b}{L} \Big) q^2 + O(q^3).
\ee
It is easy to get
\be
b'_\I(1) = 4 q^2 \Big( 1 - \f{\pi\ell}{L} \cot \f{\pi\ell}{L} \Big),
\ee
which is the same as $a'_\I(1)$ in (\ref{e42}).
We have the long interval entanglement entropy
\be \label{e47}
S_\lo(L-\ell) = \f{c}{6} \log \Big( \f{L}{\pi\e}\sin\f{\pi\ell}{L} \Big)
          + S(L)
          + 4 q^2 \Big(  1 - \f{\pi\ell}{L} \cot \f{\pi\ell}{L} \Big) + b'_\II(1) + b'_\III(1) + b'_\IV(1) + O(q^3),
\ee
with the definitions $b_\II(n)$, $b_\III(n)$, $b_\IV(n)$ in (\ref{z7}).
We cannot evaluate $b'_\II(1)$, $b'_\III(1)$, or $b'_\IV(1)$ for general $\ell$, but we can expand them by $1/c$ and $\ell$. The $c^0$ part of $b_\II(n)$ is of order $\ell^8$, the $c^0$ part of $b_\III(n)$ is of order $\ell^6$, and the $c^0$ part of $b_\IV(n)$ is of order $\ell^4$. Explicitly, we have
\be
b_\IV(n) = - \Big( \f{\pi\ell}{L} \Big)^4 \sum_{k=1}^{n-1}\f{q^{2k}}{\sin^4\f{\pi k}{n}} + O(1/c,\ell^5,(n-1)^2).
\ee
Using (\ref{z8}), we arrive at
\be
b'_\IV(1) = - \f{32\pi^5\b\ell^4}{15L^5}\Big( \f{\b^2}{L^2}+1 \Big)\Big( \f{4\b^2}{L^2}+1 \Big)q^2 + O(q^3,1/c,\ell^5).
\ee
Finally, we get
\bea \label{e68}
&& S_\lo(L-\ell) = \f{c}{6} \log \Big( \f{L}{\pi\e}\sin\f{\pi\ell}{L} \Big)
          + S(L)
          + 4 q^2 \Big(  1 - \f{\pi\ell}{L} \cot \f{\pi\ell}{L} \Big) \nn\\
&& \phantom{S_\lo(L-\ell) =}
          - \f{32\pi^5\b\ell^4}{15L^5}\Big( \f{\b^2}{L^2}+1 \Big)\Big( \f{4\b^2}{L^2}+1 \Big)q^2
          + O(q^3,1/c,\ell^5).
\eea

It is easy to see that at the leading order the entanglement entropies of the long interval and short interval are the same, as expected. The thermal entropy is not vanishing, but at the next-to-leading order in expansion of $1/c$. Different from the high temperature case, we can not ignore the exponentially suppressed terms in the low temperature. Actually in the entanglement entropies and the thermal entropy, the next-to-leading terms appear as the powers of $q$.

\begin{figure}[htbp]
  \centering
  \subfigure[$\b/L=2$]{\includegraphics[height=29mm]{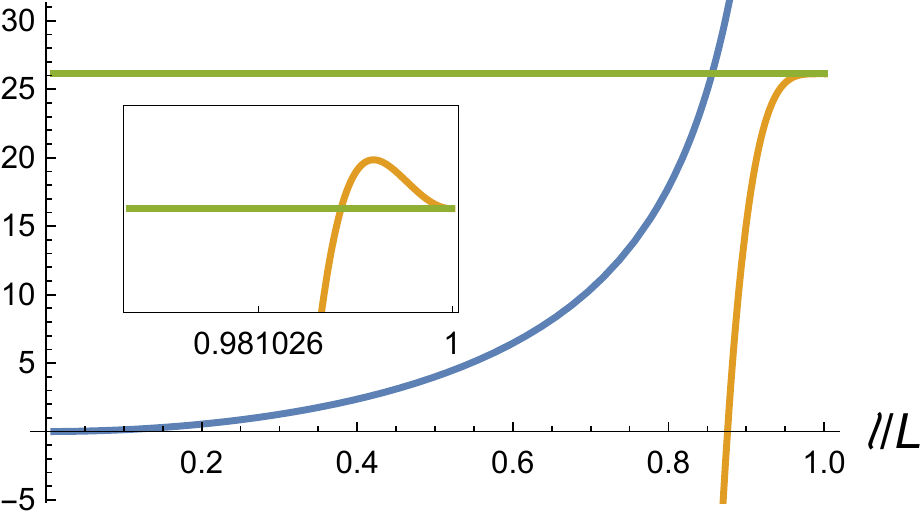}}
  \subfigure[$\b/L=3$]{\includegraphics[height=29mm]{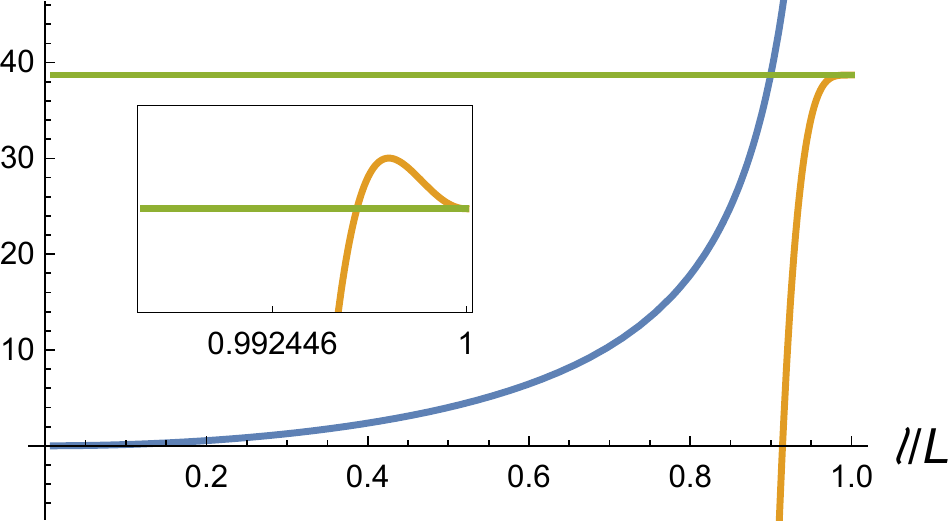}}
  \subfigure[$\b/L=5$]{\includegraphics[height=29mm]{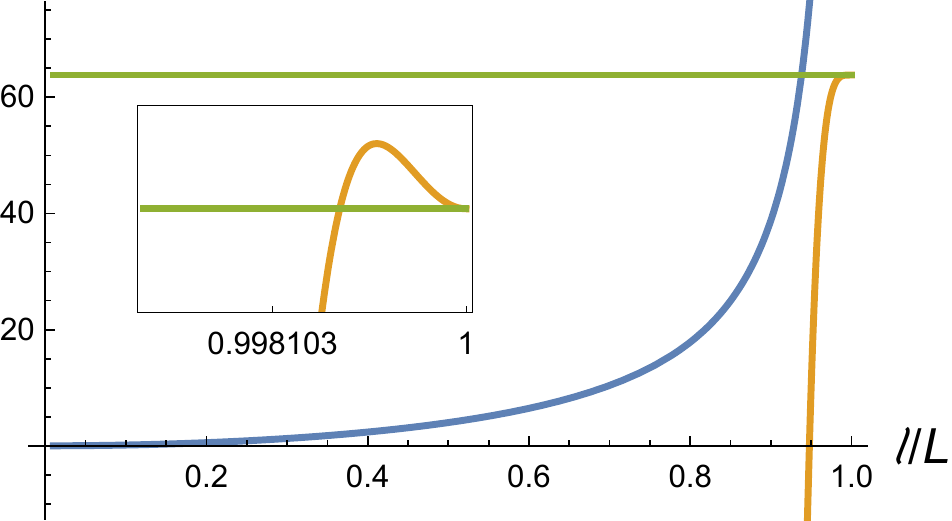}}\\
  \subfigure[$\b/L=7$]{\includegraphics[height=29mm]{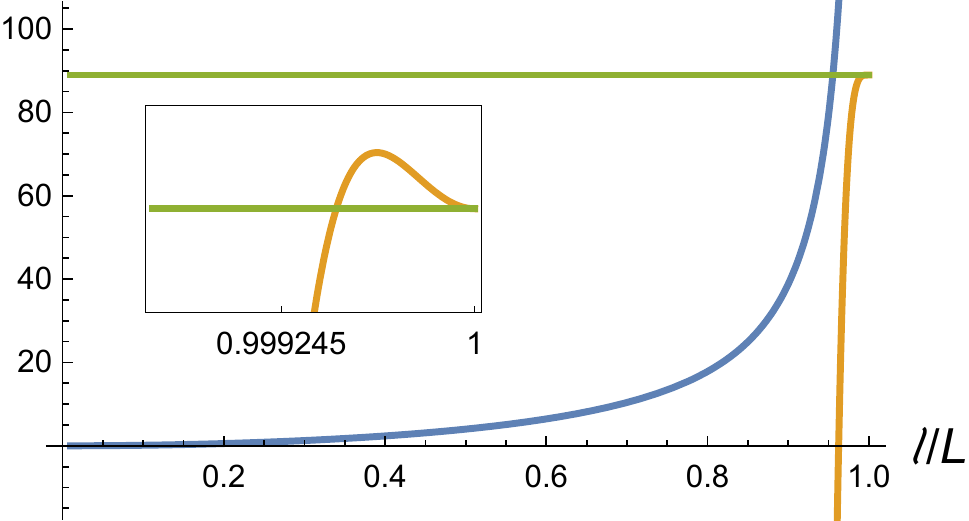}}
  \subfigure[$\b/L=9$]{\includegraphics[height=29mm]{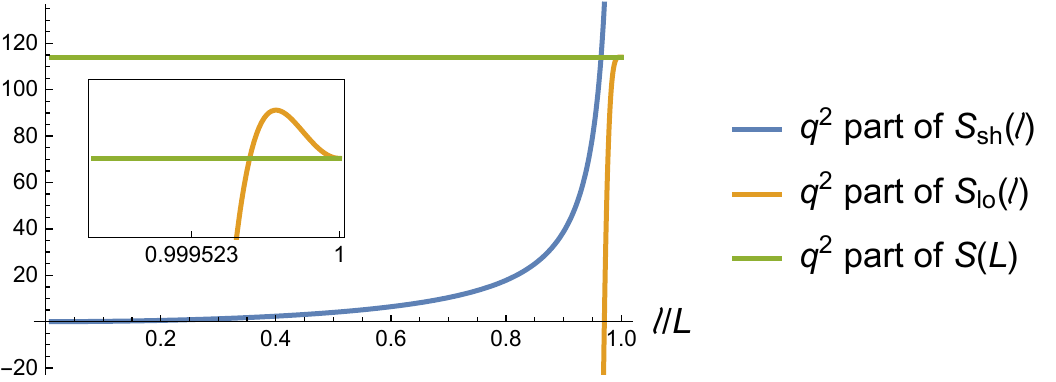}}
  \caption{The $q^2$ parts of thermal entropy and the entanglement entropies of the short interval and the long interval. There is no overlapping for the validity regions of the short interval and long interval results.}\label{slt}
\end{figure}

Let us focus on the dominant $q^2$ terms in the entropies.
 For the short interval entropy $S_\sh(\ell)$, the coefficient before $q^2$ is
\be \label{j1}
4 \Big(  1 - \f{\pi\ell}{L} \cot \f{\pi\ell}{L} \Big).
\ee
For the long interval entanglement entropy $S_\lo(\ell)=S_\lo(L-\ell)|_{\ell \to L-\ell}$ the coefficient is
\be \label{j2}
1 + \f{4\pi \b}{L}
+ 4 \Big(  1 - \f{\pi(L-\ell)}{L} \cot \f{\pi(L-\ell)}{L} \Big)
- \f{32\pi^5\b(L-\ell)^4}{15L^5}\Big( \f{\b^2}{L^2}+1 \Big)\Big( \f{4\b^2}{L^2}+1 \Big),
\ee
and for the thermal entropy $S(L)$ the coefficient is
\be \label{j3}
1 + \f{4\pi \b}{L}.
\ee
Note that in (\ref{j1}) we have omitted the possible terms of order $O(1/c,\ell^6)$, and in (\ref{j2}) we have omitted the terms of order $O(1/c,(L-\ell)^5)$. In the large $c$ limit, the omission of order $O(1/c)$ terms is justified. However, the omission of $O(\ell^6)$ or $O(L-\ell)^5$ terms would potentially spoil the validity of the results for general $\ell$.
We show the $q^2$ parts of the entropies in Figure \ref{slt}.

\subsection{Corrections to entanglement plateau}\label{sec3.5}

For small $\ell$,  we have the short interval entanglement entropy $S_\sh(\ell)$ (\ref{e67}) and the long interval entanglement entropy $S_\lo(L-\ell)$ (\ref{e68}).
We refer to the quantity $\D S=S(L)-S_\lo(L-\ell)+S_\sh(\ell)$ loosely as entanglement plateau even in the low temperature case, and we can read the correction to the entanglement plateau from the holomorphic stress tensor
\be \label{z11}
\D S = S(L)-S_\lo(L-\ell)+S_\sh(\ell) = c'_\II(1) + c'_\III(1) + c'_\IV(1) + c'_\V(1) + O(q^3),
\ee
with
\bea
&& \hspace{-5mm} c_\II(n) = \sum_{k=2}^{n-1} \Big\{ q^{2k} \sum_{0\leq j_1 < \cdots <j_{k} \leq n-1}
                    \Big( \f{\sin\f{\pi\ell}{L}}{n\sin\f{\pi\ell}{n L}} \Big)^{4k}\nn\\
&& \hspace{-5mm} \phantom{ c_\II(n) =}
                    \times \Big\lag 1 -
                    \f{(2\ii\sin\f{\pi\ell}{n L})^{4k}}{\a_T^k}\prod_{a=1}^k
                             \Big[ \ep^{\f{4\pi\ii}{n}(2j_a+\f{\ell}{L})}
                                   T(\ep^{\f{2\pi\ii}{n}(j_a+\f{\ell}{L})})
                                   T(\ep^{\f{2\pi\ii}{n}j_a}) \Big]
                    \Big\rag_\mC \Big\}, \nn\\
&& \hspace{-5mm} c_\III(n) = - \f{n(n^2-1)}{12} \sum_{k=2}^{n-1} \Big\{ q^{2k}
     \sum_{1\leq j_1< \cdots<j_{k-1}\leq n-1}
     \Big( \f{\sin\f{\pi\ell}{L}}{n\sin\f{\pi\ell}{n L}} \Big)^{4(k-1)}
     \Big(\f{2\ii}{n}\sin\f{\pi\ell}{L}\Big)^4
     \Big\lag
     \Big[  \ep^{\f{4\pi\ii\ell}{nL}}T(\ep^{\f{2\pi\ii\ell}{nL}}) + T(0)  \Big]\nn\\
&& \hspace{-5mm} \phantom{c_\III(n)=} \times
     \f{(2\ii\sin\f{\pi\ell}{n L})^{4(k-1)}}{\a_T^{k-1}}\prod_{a=1}^{k-1}
                             \Big[ \ep^{\f{4\pi\ii}{n}(2j_a+\f{\ell}{L})}
                                   T(\ep^{\f{2\pi\ii}{n}(j_a+\f{\ell}{L})})
                                   T(\ep^{\f{2\pi\ii}{n}j_a}) \Big]
     \Big\rag_\mC \Big\}, \nn\\
&& \hspace{-5mm} c_\IV(n) = \sum_{k=1}^{n-1} \Big\{ q^{2k} \sum_{0\leq j_1< \cdots<j_k\leq n-1}
            \f{(\f{2\ii}{n}\sin\f{\pi\ell}{L})^{4k}}{\a_T^k} \Big \lag
            \prod_{a=1}^k \Big[
                \ep^{\f{4\pi\ii}{n}(2j_a+1-\f{\ell}{L})}
                T(\ep^{\f{2\pi\ii}{n}(j_a+1-\f{\ell}{L})})
                T(\ep^{\f{2\pi\ii}{n}j_a})
              \Big]
\Big\rag_\mC \Big\}, \nn\\
&& \hspace{-5mm} c_\IV(n) =  \f{n(n^2-1)}{12} \sum_{k=2}^{n-1} \Big\{  q^{2k} \sum_{1\leq j_1< \cdots<j_{k-1}\leq n-1} \f{(\f{2\ii}{n}\sin\f{\pi\ell}{L})^{4k}}{\a_T^{k-1}} \Big \lag
   \Big[  \ep^{\f{4\pi\ii}{n}}T(\ep^{\f{2\pi\ii}{n}}) + \ep^{\f{4\pi\ii\ell}{n L}}T(\ep^{\f{2\pi\ii\ell}{n L}})  \Big]\nn\\
&& \hspace{-5mm} \phantom{c_\IV(n) =}
\times\prod_{a=1}^{k-1} \Big[
                \ep^{\f{4\pi\ii}{n}(2j_a+1+\f{\ell}{L})}
                T(\ep^{\f{2\pi\ii}{n}(j_a+1)})
                T(\ep^{\f{2\pi\ii}{n}(j_a + \f{\ell}{L})})
              \Big]\Big\rag_\mC \Big\}.
\eea

We expand the result (\ref{z11}) by small $\ell$ while keeping the central charge $c$ general. It is easy to see that $c_\II(n)$, $c_\IV(n)$ are of order $\ell^4$, and $c_\III(n)$, $c_\V(n)$ are of order $\ell^6$. Explicitly, we get
\bea
&& c_\II(n) = -\f{4}{c} \Big( \f{\pi\ell}{L} \Big)^4 \sum_{k=2}^{n-1}C_{n-2}^{k-2}q^{2k}\sum_{j=1}^{n-1} \f{1}{\sin^4\f{\pi j}{n}} + O(\ell^5,(n-1)^2), \nn\\
&& c_\IV(n) = \Big( \f{\pi\ell}{L} \Big)^4 \sum_{k=1}^{n-1} \f{q^{2 k}}{\sin^4\f{\pi k}{n}} + O(\ell^5,(n-1)^2).
\eea
Then we get the corrections to the entanglement plateau
\be \label{z12}
S(L)-S_\lo(L-\ell)+S_\sh(\ell) = \f{32q^2}{15}
                                 \Big( \f{\pi\ell}{L} \Big)^4
                                 \Big[ \f{1}{c} +  \f{\pi \b}{L} \Big( \f{\b^2}{L^2}+1 \Big)\Big( \f{4\b^2}{L^2}+1 \Big) \Big]
                                 + O(q^3,\ell^5).
\ee
Note that in the above result we do not require the central charge to be large, but we have only incorporated the contributions from the vacuum module.

\subsection{Low temperature case with nonvacuum module} \label{sec4}


In this subsection we consider the low temperature case with the leading contributions from a holomorphic nonvacuum module.
We consider the module with a general holomorphic primary operator $\mX$ of conformal weight $h_\mX$ and normalization $\a_\mX$.
It was shown in \cite{Cardy:2014jwa} the leading order correction to the single-interval entanglement entropy from the module $\mX$ takes a universal form
\be \label{e92}
\d_\mX S(\ell) = 2 h_\mX q^{h_\mX} \Big( 1 - \f{\pi\ell}{L} \cot \f{\pi\ell}{L} \Big) + O(q^{h_\mX+1},q^{2h_\mX}).
\ee
It was believed that this applies to a general interval as long as the length $\ell$ cannot be comparable to length of the system $L$. 

Due to the presence of the primary module, we find that the corrections to the density matrix and the reduced density matrices are respectively
\be \label{e79}
\d_\mX \r = \f{q^{h_\mX}}{\a_\mX} |\mX\rag\lag\mX| + O(q^{h_\mX+1}), ~~
\d_\mX \r_A = \f{q^{h_\mX}}{\a_\mX} \r_{A,\mX} + O(q^{h_\mX+1}).
\ee
Using the same method as in subsection~\ref{sec3.2} we may get the corrections to the short interval entanglement entropy
\be \label{z9}
\d_\mX S_\sh(\ell) = 2 h_\mX q^{h_\mX} \Big(  1 - \f{\pi\ell}{L} \cot \f{\pi\ell}{L} \Big)
                   + \p_n a^\II_\mX(n)|_{n=1}+O(q^{h_\mX+1},q^{2 h_\mX}),
\ee
with
\bea
&& a_\mX^\II(n) = -\sum_{k=2}^n \Big\{ q^{kh_\mX} \Big( \f{\sin\f{\pi\ell}{L}}{n\sin\f{\pi\ell}{n L}} \Big)^{2kh_\mX}
                   \Big( 1+q^{h_\mX} \Big(\f{\sin\f{\pi\ell}{L}}{n\sin\f{\pi\ell}{n L}}\Big)^{2h_\mX} \Big)^{n-k} \\
&& \phantom{g_\mX^\II(n) =} \times
                   \sum_{\mZ_1,\cdots,\mZ_k}\sum_{0\leq j_1<\cdots<j_k\leq n-1}
                   \Big\lag
                     \prod_{a=1}^k \Big[ D_{\mX\mX\mZ_a} \Big( \ep^{\f{2\pi\ii}{n}j_a} \big( \ep^{\f{2\pi\ii\ell}{n L}} -1\big) \Big)^{h_{\mZ_a}} \mZ_a(\ep^{\f{2\pi\ii}{n}j_a}) \Big]
                   \Big\rag_\mC \Big\}. \nn
\eea
In $a_\mX^\II(n)$ the quantities $D_{\mX\mX\mZ_a}$ are defined by the OPE of $\mX(z_1)\mX(z_2)$
\bea
&& \mF_\mX(z_1,z_2) = 1+ \sum_\mY \f{C_{\mX\mX\mY}}{\a_\mX\a_\mY} \sum_{r=0}^\inf \f{a_\mY^r}{r!}(z_1-z_2)^{h_\mY+r}\p^r\mY(z_2) \nn\\
&& \phantom{\mF_\mX(z,w)}
                = 1+\sum_\mZ D_{\mX\mX\mZ} (z_1-z_2)^{h_\mZ} \mZ(z_2),
\eea
with $a_\mY^r = \f{C_{h_\mY+r-1}^r}{C_{2h_\mY+r-1}^r}$.
The summation for $\mY$ runs over all the nonidentity holomorphic quasiprimary operators with each $\mY$ being of conformal weight $h_\mY$, and the summation for $\mZ$ runs over all the nonidentity holomorphic operators, including the quasiprimary operators and their derivatives. It is possible that the term $\p_n a^\II_\mX(n)|_{n=1}$ give the same order of contribution as $q^{h_\mX}$. It would be nice if $\p_n a^\II_\mX(n)|_{n=1}$ can be evaluated without taking small $\ell$ expansion.


Similarly, we can read the correction to the thermal entropy and the entanglement entropy of the long interval. The correction to the thermal entropy from the primary module $\mX$ is
\be \label{e80}
\d_\mX S(L) = \Big( 1 + \f{2\pi h_\mX \b}{L} \Big) q^{h_\mX} + O(q^{h_\mX+1},q^{2h_\mX}).
\ee
The corrections to the long interval entanglement entropy is
\be \label{z10}
\d_\mX S_\lo(L-\ell) = \d_\mX S(L)
          + 2 h_\mX q^{h_\mX} \Big(  1 - \f{\pi\ell}{L} \cot \f{\pi\ell}{L} \Big)
          + \p_n b^\II_\mX(n)|_{n=1}+ \p_n b^\III_\mX(n)|_{n=1} +O(q^{h_\mX+1},q^{2h_\mX}),
\ee
with the definitions
\bea
&& \hspace{-7mm}
   b^\II_\mX(n)=
    -q^{n h_\mX} \Big( \f{\sin\f{\pi\ell}{L}}{n\sin\f{\pi\ell}{n L}} \Big)^{2 n h_\mX} \nn\\
&& \hspace{-7mm}\phantom{b^\II_\mX(n)=} \times
    \sum_{k=2}^n \sum_{\mZ_1,\cdots,\mZ_k}\sum_{0\leq j_1<\cdots<j_k\leq n-1}
      \Big\lag
         \prod_{a=1}^k \Big[ D_{\mX\mX\mZ_a} \Big( \ep^{\f{2\pi\ii}{n}j_a} \big( \ep^{\f{2\pi\ii\ell}{n L}} - 1 \big) \Big)^{h_{\mZ_a}} \mZ_a(\ep^{\f{2\pi\ii}{n}j_a}) \Big]
      \Big\rag_\mC,  \\
&& \hspace{-7mm}b^\III_\mX(n) = -\sum_{k=1}^{n-1} q^{k h_\mX} \sum_{0\leq j_1< \cdots<j_k\leq n-1}
\f{(\f{2\ii}{n}\sin\f{\pi\ell}{L})^{2k h_\mX}}{\a_\mX^k}
\Big\lag \prod_{a=1}^k \Big[
                \ep^{\f{2\pi \ii h_\mX}{n}(2j_a+1-\f{\ell}{L})}
                \mX(\ep^{\f{2\pi\ii}{n}(j_a+1-\f{\ell}{L})})
                \mX(\ep^{\f{2\pi\ii}{n}j_a})
              \Big]
\Big\rag_\mC. \nn
\eea

The long interval result (\ref{z10}) is not universal and depends on the structure constants, and the short interval result (\ref{z9}) is also possibly not universal. One can compare (\ref{z9}), (\ref{z10}) with the result (\ref{e92}), which was obtained in \cite{Cardy:2014jwa}. We get the different results by using a refined $n \to 1$ limit. We sum all the terms of orders $q^{h_\mX}$, $q^{2 h_\mX}$, $\cdots$, $q^{(n-1)h_\mX}$, $q^{n h_\mX}$ before taking the $n \to 1$ limit, while in \cite{Cardy:2014jwa} only the term of order $q^{h_\mX}$ was kept in obtaining (\ref{e92}). Though it is fine to keep only the order $q^{h_\mX}$ term in calculating the $n$-th R\'enyi entropy with $n=2,3,4,\cdots$, we need to keep all the terms of orders $q^{h_\mX}$, $q^{2 h_\mX}$, $\cdots$, $q^{(n-1)h_\mX}$, $q^{n h_\mX}$  to get the correct $n\to 1$ limit. One justification for our treatment is that $\d_\mX S(\ell)$ in (\ref{e92}) is ill-defined in the limit $\ell \to L$ while $\d_\mX S_\lo(L-\ell)$ in (\ref{z10}) is well-defined in the limit $\ell \to 0$. Note that it is still possible (\ref{e92}) is correct for a short interval, i.e., that in (\ref{z9}) it is possible $\p_n a^\II_\mX(n)|_{n=1} \sim O(q^{2h_\mX})$.

Summing up all the contributions, we find the correction to the entanglement plateau
\be
\d_\mX ( S(L) - S_\lo(L-\ell) + S_\sh(\ell) ) = \p_n c^\II_\mX(n)|_{n=1}+ \p_n c^\III_\mX(n)|_{n=1} +O(q^{h_\mX+1},q^{2h_\mX}),
\ee
with the definitions
\bea \label{z13}
&& \hspace{-7mm} c^\II_\mX(n)=
    - \sum_{k=2}^{n-1}  \Big\{ q^{k h_\mX} \Big( \f{\sin\f{\pi\ell}{L}}{n\sin\f{\pi\ell}{n L}} \Big)^{2 k h_\mX}
    \Big[ \Big( 1+ q^{h_\mX} \Big( \f{\sin\f{\pi\ell}{L}}{n\sin\f{\pi\ell}{n L}} \Big)^{2h_\mX} \Big)^{n-k}
          - q^{(n-k)h_\mX} \Big( \f{\sin\f{\pi\ell}{L}}{n\sin\f{\pi\ell}{n L}} \Big)^{2(n-k)h_\mX}  \Big] \nn\\
&& \hspace{-7mm} \phantom{b^\II_\mX(n)=} \times
    \sum_{\mZ_1,\cdots,\mZ_k}\sum_{0\leq j_1<\cdots<j_k\leq n-1}
      \Big\lag
         \prod_{a=1}^k \Big[ D_{\mX\mX\mZ_a} \Big( \ep^{\f{2\pi\ii}{n}j_a} \big( \ep^{\f{2\pi\ii\ell}{n L}} - 1 \big) \Big)^{h_{\mZ_a}} \mZ_a(\ep^{\f{2\pi\ii}{n}j_a}) \Big]
      \Big\rag_\mC \Big\},  \\
&& \hspace{-7mm} c^\III_\mX(n) = \sum_{k=1}^{n-1} q^{k h_\mX} \sum_{0\leq j_1< \cdots<j_k\leq n-1}
\f{(\f{2\ii}{n}\sin\f{\pi\ell}{L})^{2k h_\mX}}{\a_\mX^k}
\Big\lag \prod_{a=1}^k \Big[
                \ep^{\f{2\pi \ii h_\mX}{n}(2j_a+1+\f{\ell}{L})}
                \mX(\ep^{\f{2\pi\ii}{n}(j_a+1)})
                \mX(\ep^{\f{2\pi\ii}{n}(j_a+\f{\ell}{L})})
              \Big]
\Big\rag_\mC. \nn
\eea
We expand the results (\ref{z13}) by small $\ell$ and get
\bea
&& c_\mX^\II(n) = \f{q^{h_\mX}}{2} \sum_{\mY} \Big[  \f{C_{\mX\mX\mY}^2}{\a_\mX^2\a_\mY} \Big( \f{\pi\ell}{L} \Big)^{2h_\mY}
                  \sum_{j=1}^{n-1} \f{1}{(\sin\f{\pi j}{n})^{2h_\mY}} + O( \ell^{2h_\mY+1}, \ell^{3h_\mY}) \Big]
                  + O(q^{2 h_\mX},(n-1)^2), \nn\\
&& c_\mX^\III(n) = \Big( \f{\pi\ell}{L} \Big)^{2h_\mX} \sum_{k=1}^{n-1} \f{q^{k h_\mX}}{(\sin\f{\pi k}{n})^{2h_\mX}} + O(\ell^{2h_\mX+1},(n-1)^2).
\eea
The summation for $\mY$ runs over all the nonidentity holomorphic quasiprimary operators.
Finally we find
\bea \label{z14}
&& \hspace{-8mm}
   \d_\mX ( S(L) - S_\lo(L-\ell) + S_\sh(\ell) ) =
   \f{\sqrt{\pi}q^{h_\mX}}{4} \sum_\mY \Big[ \f{C_{\mX\mX\mY}^2}{\a_\mX^2\a_\mY} \f{\G(h_\mY+1)}{\G(h_\mY+3/2)} \Big( \f{\pi\ell}{L} \Big)^{2h_\mY} + O( \ell^{2h_\mY+1}, \ell^{3h_\mY}) \Big] \nn\\
&& \hspace{-8mm} ~~~~~~~~~
   + \Big( \f{\pi\ell}{L} \Big)^{2h_\mX} \p_n \Big[ \sum_{k=1}^{n-1} \f{q^{k h_\mX}}{(\sin\f{\pi k}{n})^{2h_\mX}} \Big]\Big|_{n \to 1}
   + O(q^{h_\mX+1},q^{2h_\mX},\ell^{2h_\mX+1}).
\eea
In the summation over $\mY$,  the  holomorphic quasiprimary operator with the smallest conformal weight in each module dominates. For the vacuum module it is the stress tensor $T$, and for a nonvacuum module it is just the primary operator.
The summation of $\mY$ in (\ref{z14}) runs over $T$ and all the nonidentity holomorphic primary operators. For the stress tensor, it gives the correction (\ref{z12}). Note that that the result (\ref{z14}) does not take a universal form, as it depends on the structure constants.

\section{Conclusion and discussions}\label{sec5}

In this work, we studied the single-interval entanglement entropies at finite temperature in 2D CFT. We focused on the high temperature case with $\b \ll L$ and the low temperature case with $L\ll \b $. In particular we computed the entanglement entropies in the short and large interval limits. This allows us to discuss the subleading correction to the entanglement plateau
\be
\D S=S(L)-|S(L-\ell)-S(\ell)|
\ee
where $S(L)$ is the thermal entropy of the system at the finite temperature. A general lesson is that the Araki-Lieb inequality is robust and cannot be saturated for a finite $\ell$ if the next-to-leading order contributions of large $c$ limit are taken into account.

For the large $c$ CFT with a gravity dual, it was found that there could be a holographic entanglement plateau at a high temperature. As suggested in \cite{Faulkner:2013ana}, $\D S$ is the mutual information between the interior of the black hole and the region enclosed by the minimal surface $\g_{\ell}$ \cite{Faulkner:2013ana}. We explicitly computed this mutual information in this work. In the semi-classical AdS$_3$/CFT$_2$, we showed that $\D S$ is nonvanishing but is always an order $c^0$ effect, including both the contributions from the vacuum module and other primary modules.

In computing the entanglement entropies at a high temperature, we omitted the finite size effect, which contributes the exponentially suppressed terms. This simplifies the computation significantly and allows us to relate the computation with the computation on the R\'enyi entropy of the two disconnected intervals on a complex plane. One important consequence is that the leading contribution from the a nonvacuum module takes a universal form.

On the other hand, in the low temperature case we cannot ignore the exponentially suppressed terms.
We used two different approaches to compute the thermal corrections to the entanglement entropies and found consistent results.
Quite interestingly we found that the leading thermal correction to the long interval entanglement entropy actually does not take a universal form. 
Instead, the leading correction to entanglement plateau actually depends on the details of the theory.



It is remarkable that our treatment in this work does not restrict to the large $c$ CFT, and can be applied to a general CFT. In 2D CFT, the vacuum module is special as it includes the stress tensor which encodes the information on the central charge. Therefore the discussion on the vacuum module in this work certainly applies to other CFTs. At the high
temperature the case is related to the double interval mutual information on the complex plane. In the latter case the leading contribution of the nonvacuum module to the mutual information is of universal form. At the low temperature, the picture is similar but the leading contributions from nonvacuum modules depend on the details of the theory.
Simply speaking, in a general 2D CFT the corrections from both vacuum and nonvacuum modules are not suppressed by $1/c$ and so the entanglement plateau disappears.


\section*{Acknowledgements}

We thank Peng-xiang Hao for a partial collaboration on this work, especially the calculation in appendix~\ref{appC}.
We thank Erik Tonni for helpful discussions. We thank the Galileo Galilei Institute for
Theoretical Physics for the hospitality and the INFN for partial support during the completion of this work. BC would like to thank Centro de Ciencias de Benasque Pedro Pascual for hospitality during the completion of this work.
BC was in part supported by NSFC Grant No.~11275010, No.~11335012 and No.~11325522.
ZL was supported by NSFC Grant No.~11575202.
JJZ was supported by the ERC Starting Grant 637844-HBQFTNCER and in part by Italian Ministero dell'Istruzione, Universit\`a e Ricerca (MIUR) and Istituto Nazionale di Fisica Nucleare (INFN) through the ``Gauge Theories, Strings, Supergravity'' (GSS) research project.

\appendix

\section{Mutual information of two intervals on a complex plane}\label{appA}

Here we review the useful property of the mutual information $I(x)$ between two intervals in a large $c$ CFT with contributions from the vacuum module \cite{Headrick:2010zt,Calabrese:2010he,Barrella:2013wja,Chen:2013kpa,Chen:2013dxa,Beccaria:2014lqa,Li:2016pwu}. The mutual information can be organized by orders of $c$
\be
I(x) = I_\cL(x) + I_\NL(x) +\cdots,
\ee
where $x$ is the cross ratio.
The leading part of the mutual information is universal and do not depend on the details of the CFT
\be \label{miL}
I_\cL(x) = \lt\{ \ba{ll}
0 & ~\when~ x<1/2 \\
\df{c}{3} \log\df{x}{1-x} & ~\when~ x>1/2,
\ea\rt.
\ee
and with contributions of only the vacuum module the next-to-leading part can be written in expansion of small $x$
\be \label{z92}
I_\NL(x) = \frac{x^4}{630}+\frac{2 x^5}{693}+\frac{15 x^6}{4004}+\frac{x^7}{234}+\frac{167 x^8}{36036}+\frac{69422 x^9}{14549535}+\frac{122x^{10}}{24871}+O(x^{11}).
\ee
One has \cite{Headrick:2010zt}
\be
I_\NL(x) = I_\NL(1-x).
\ee
For a nonvacuum module with a primary operator $\mX$ of the scaling dimension $\D_\mX$, there is a universal correction at the leading order\cite{Calabrese:2010he}
\be \label{e91}
\d_\mX I(x) =   \frac{\sqrt{\pi} \Gamma(2 \Delta_\mX +1) x^{2 \Delta_\mX }}{4^{2 \Delta_\mX +1}\Gamma(2 \Delta_\mX  +3/2)} + O(x^{2\D_\mX+1},x^{3\D_\mX}).
\ee

In fact, for any 2D CFT the small $x$ expansion of the mutual information can be written as \cite{Beccaria:2014lqa,Li:2016pwu}
\be
I(x) = \lim_{n \to 1} \f{1}{n-1}\sum_K \a_K d_K^2 x^{h_K+\bar h_K} {}_2F_1(h_K,h_K;2h_K;x) {}_2F_1(\bar h_K,\bar h_K;2\bar h_K;x),
\ee
where the summation $K$ runs over all orthogonalized quasiprimary operators $\Phi_K$, with conformal weights $(h_K,\bar h_K)$ and normalization $\a_K$, in the $n$-fold CFT that we call $\CFT^n$, and $d_K$ is the OPE coefficient of twist operators. It is just (\ref{z92}) with the contributions from only the vacuum module. The leading contribution from a nonvacuum module takes the universal form (\ref{e91}), while the subleading contributions are not universal and depend on details of the CFT. Also the subleading contributions from different modules are mixed and cannot be separated.


\section{Analytical continuation}\label{appB}

In the appendix, we prove the following identity\footnote{We thank Peng-xiang Hao for his contributions to this appendix.}
\be \label{z8}
\p_n \Big( \sum_{k=1}^{n-1}\f{q^{2k}}{\sin^4\f{\pi k}{n}} \Big)\Big|_{n=1}
= \f{32\pi\b}{15L}\Big( \f{\b^2}{L^2}+1 \Big)\Big( \f{4\b^2}{L^2}+1 \Big)q^2 + O(q^3).
\ee
Note that $q=\ep^{-2\pi\b/L}$.

We consider the Mellin transform and its inverse transform
\bea
&& F(s) = \int_0^\inf f(x) x^{s-1} d x, \nn\\
&& f(x) = \f{1}{2\pi\ii} \int_{c-\ii\inf}^{c+\ii\inf} F(s) x^{-s} ds.
\eea
We choose
\be
F(s) = \f{1}{[\sin(\pi s)]^4},
\ee
and so
\be
f(x) = \f{(\log x)^3 + 4\pi^2 \log x}{6\pi^4(x-1)}.
\ee
Then we get
\be
\sum_{k=1}^{n-1}\f{q^{2k}}{\sin^4\f{\pi k}{n}} = \int_0^\inf \f{q^2 x^{1/n}-q^{2n}x}{x(1-q^2 x^{1/n})} f(x) d x,
\ee
which leads to
\be
\p_n \Big( \sum_{k=1}^{n-1}\f{q^{2k}}{\sin^4\f{\pi k}{n}} \Big)\Big|_{n=1}
= \int_0^\inf \f{q^2 [ \log(q^2) + \log x ]}{q^2 x -1} f(x) d x.
\ee
The integral on the right-hand side is convergent for $\Re q^2 \leq 0$, $\Im q^2 \neq 0$, and with an analytical continuation we finally get
\be
\p_n \Big( \sum_{k=1}^{n-1}\f{q^{2k}}{\sin^4\f{\pi k}{n}} \Big)\Big|_{n=1} =
\f{q^2\log(q^2)[\log(q^2)+4\pi^2][\log(q^2)+16\pi^2]}{120\pi^4(q^2-1)}.
\ee
An immediate check of the result is that when $q^2 \to 1$ it is $\f{8}{15}$, and this is consistent with (\ref{z83}). Then we get (\ref{z8}).

\section{Low temperature case from method of twist operators} \label{appC}

As a double check of the results in sections~\ref{sec3} and \ref{sec4}, we calculate the short and long interval entanglement entropies using the OPE of the twist operators in this section.

The replica trick leads to an $n$-fold CFT, which we call $\CFT^n$, on a nontrivial Riemann surface. The partition function of $\CFT^n$ can be computed by a correlation function of the twist operators.
The twist operators $\s$, $\td\s$ are primary operators with conformal dimension \cite{Calabrese:2004eu}
\be
h_\s=h_{\td\s}=\f{c(n^2-1)}{24n}.
\ee
When the interval $A=[0,\ell]$ is short, we may use the OPE of the twist operators \cite{Calabrese:2009ez,Headrick:2010zt,Calabrese:2010he,Chen:2013kpa}
\be \label{opet}
\s(\ell)\td \s(0) = \Big(\f{\e}{\ell}\Big)^{2h_\s} \sum_K d_K \sum_{r\geq0} \f{a_K^r}{r!} \ell^{h_K+r} \p^r \Phi_K(0), ~~
a_K^r \equiv \f{C_{h_K+r-1}^r}{C_{2h_K+r-1}^r},
\ee
with the summation $K$ being over all CFT$^n$ quasiprimary operators $\Phi_K$. To the order we consider in this paper, we only need $\Phi_K$ that can be written as a direct product of the quasiprimary operators in different replicas
\be
\Phi_K^{j_1j_2\cdots j_k} = \phi_1^{j_1}\phi_2^{j_2}\cdots\phi_k^{j_k},
\ee
where $0\leq j_i\leq n-1$ labels the replica.
From the OPE coefficients $d_K^{j_1j_2\cdots j_k}$ for $\Phi_K^{j_1j_2\cdots j_k}$, we may define
\be
b_K = \sum_{j_1,j_2,\cdots,j_k}d_K^{j_1j_2\cdots j_k}, ~~
a_K = - \lim_{n \to 1} \f{b_K}{n-1}.
\ee

For the vacuum module, we only need the quasiprimary operators $T_j$, $\mA_j$, $T_{j_1}T_{j_2}$, with $0 \leq j\leq n-1$, $0 \leq j_1 < j_2 \leq n-1$, the corresponding $d_K$ can be found in \cite{Calabrese:2010he,Chen:2013kpa,Chen:2013dxa}, and the corresponding $b_K$ can be found in \cite{Chen:2016lbu,He:2017vyf}, from which we may get
\be \label{e72}
a_T = -\frac{1}{6}, ~~
a_\mA =0, ~~
a_{TT} = -\frac{1}{30 c}.
\ee
For a quasiprimary operator $\mY$, we may have the CFT$^n$ quasiprimary operators $\mY_{j_1}\mY_{j_2}$ with $0 \leq j_1 < j_2 \leq n-1$. The corresponding $d_K$ is \cite{Calabrese:2010he,Perlmutter:2013paa,Chen:2014kja}
\be \label{e75}
d_{\mY\mY}^{j_1j_2} = \f{\ii^{2h_\mY}}{2^{2h_\mY}\a_\mY}\f{1}{[\sin\f{\pi(j_1-j_2)}{n}]^{2h_\mY}} + O(n-1),
\ee
from which we get
\bea \label{e82}
&& b_{\mY\mY} = \f{\ii^{2h_\mY}}{2^{2h_\mY+1}\a_\mY} \sum_{j=1}^{n-1} \f{1}{(\sin\f{\pi j}{n})^{2h_\mY}} + O(n-1)^2, \nn\\
&& a_{\mY\mY} = - \f{\ii^{2h_\mY}\sqrt{\pi}\G(h_\mY+1)}{2^{2(h_\mY+1)}\a_\mY\G(h_\mY+3/2)}.
\eea
Note that we have used the relation\cite{Calabrese:2010he}
\be \label{z83}
\p_n \Big[ \sum_{j=1}^{n-1} \f{1}{(\sin\f{\pi j}{n})^{2h_\mY}} \Big]\Big|_{n=1} = \f{\sqrt{\pi}\G(h_\mY+1)}{2\G(h_\mY+3/2)}.
\ee

In the following calculation we need the one-point function of operator $\mY$ on a vertical cylinder capped with two operators $\mX$, $\mZ$ on the two ends, and we denote it by $\lag \mX | \mY(w) | \mZ \rag$.
When $\mX=\mZ$, we also define $\lag \mY \rag_\mX = \lag \mX | \mY | \mX \rag /\a_\mX$.
We evaluate it by mapping the cylinder to a complex plane by conformal transformation $z=\ep^{\f{2\pi\ii w}{L}}$. In the following calculation we need the relations
\bea \label{e73}
&& \lag T \rag_0 = \f{\pi^2 c}{6L^2}, ~~ \lag T \rag_T = \f{\pi^2 (c-48)}{6L^2}, \nn\\
&& \lag T | T(w) | 0 \rag = - \f{2\pi^2 c z^2}{L^2}, ~~
   \lag 0 | T(w) | T \rag = - \f{2\pi^2 c}{L^2z^2}.
\eea
Note that we use 0 to denote the identity operator 1 which corresponds to the ground state $|0\rag$. For a nonidentity primary operator $\mX$ and a nonidentity primary operator $\mY$, we have
\bea \label{e83}
&& \lag T \rag_\mX = \f{\pi^2 (c-24h_\mX)}{6L^2}, ~~
   \lag \mY \rag_0 = 0, ~~
   \lag \mY \rag_\mX = \Big( \f{2\pi\ii}{L} \Big)^{h_\mY} \f{C_{\mX\mY\mX}}{\a_\mX}, \nn\\
&& \lag \mX | \mX(w) | 0 \rag = \a_\mX \Big( \f{2\pi\ii z}{L} \Big)^{h_\mX}, ~~
   \lag 0 | \mX(w) | \mX \rag = \a_\mX \Big( \f{2\pi\ii}{L z} \Big)^{h_\mX}.
\eea
Note that for the structure constant $C_{\mX\mY\mX}$ being nonvanishing, we need that $\mY$ is bosonic, i.e., that $h_\mY$ is an integer. There is relation $C_{\mX\mY\mX} = (-)^{h_\mY} C_{\mX\mX\mY}$.

\subsection{Contributions from the vacuum module}

For a short interval $A=[0,\ell]$, we have the reduced density matrix (\ref{z15}). Using the twist operators we get
\be \label{e74}
\tr_A \r_A^n = \sum_{k=0}^n \f{q^{2k}}{\a_T^k} \sum_{0 \leq j_1 < \cdots < j_k \leq n-1}
\lag T_{j_1}\cdots T_{j_k}| \s(\ell)\td\s(0) | T_{j_1}\cdots T_{j_k} \rag + O(q^3).
\ee
Using the OPE of the twist operators we have
\bea
&&\hspace{-8mm} \f{1}{\a_T^k} \sum_{0 \leq j_1 < \cdots < j_k \leq n-1} \lag T_{j_1}\cdots T_{j_k}| \s(\ell)\td\s(0) | T_{j_1}\cdots T_{j_k} \rag = \Big( \f{\e}{\ell} \Big)^{2h_\s} C_n^k \Big[
 1 + \ell^2 b_T \f{(n-k)\lag T \rag_0 + k\lag T \rag_T}{n}  \nn\\
&&\hspace{-8mm} ~~~ + \ell^4 \Big(  b_\mA \f{(n-k)\lag \mA \rag_0 + k\lag \mA \rag_T}{n}
                     + b_{TT} \f{(n-k)(n-k-1)\lag T \rag_0^2 + 2k(n-k)\lag T \rag_0\lag T \rag_T + k(k-1)\lag T \rag_T^2  }{n(n-1)}
                \Big)\nn\\
&&\hspace{-8mm} ~~~ + O(\ell^6)
\Big],
\eea
with which we get
\bea
&&\hspace{-6mm} \tr_A\r_A^n = \Big( \f{\e}{\ell} \Big)^{2h_\s} \big\{ (1+q^2)^n
  + \ell^2 b_T (1+q^2)^{n-1} ( \lag T \rag_0 + q^2\lag T \rag_T ) 
  + \ell^4\big[  b_\mA (1+q^2)^{n-1} ( \lag \mA \rag_0 + q^2 \lag \mA \rag_T )\nn\\
&&\hspace{-6mm} \phantom{\tr_A\r_A^n =}
                + b_{TT} (1+q^2)^{n-2} (\lag T \rag_0^2
                             + 2q^2 \lag T \rag_0\lag T \rag_T
                             + q^4 \lag T \rag_T^2
                         )
  \big] + O(\ell^6)
 \big\} + O(q^3).
\eea
Then we get the short interval entanglement entropy
\bea
&& S_\sh(\ell) = \f{c}{6}\log\f{\ell}{\e} + \big[ \ell^2 a_T \lag T \rag_0 + \ell^4 a_{TT} \lag T \rag_0^2 + O(\ell^6) \big]\\
&& \phantom{S_\sh(\ell) =}
   + \big[ \ell^2 a_T ( \lag T \rag_T - \lag T \rag_0) + 2 \ell^4 a_{TT}\lag T \rag_0( \lag T \rag_T - \lag T \rag_0) + O(\ell^6) \big]q^2 + O(q^3).\nn
\eea
Using (\ref{e72}), (\ref{e73}), we further find
\be \label{e77}
S_\sh(\ell) = \f{c}{6}\log\f{\ell}{\e} + \Big( -\f{\pi^2 c \ell^2}{36L^2} -\f{\pi^4 c \ell^4}{1080L^4} + O(\ell^6) \Big)
+ \Big( \f{4\pi^2 \ell^2}{3L^2} + \f{4\pi^4 \ell^4}{45L^4} + O(\ell^6) \Big) q^2 + O(q^3).
\ee
This is consistent with (\ref{e67}) and the results in \cite{Cardy:2014jwa,Chen:2014hta,Chen:2015uia,Chen:2016lbu}.

The complement of $A$ is a long interval $B$ with length $L-\ell$. Instead of (\ref{e74}), for a long interval we have \cite{Chen:2014ehg}
\be
\tr_B \r_B^n = \sum_{k=0}^n \f{q^{2k}}{\a_T^k} \sum_{0 \leq j_1 < \cdots < j_k \leq n-1}
\lag T_{j_1}\cdots T_{j_k}| \s(\ell)\td\s(0) | T_{j_1+1}\cdots T_{j_k+1} \rag + O(q^3).
\ee
Using OPE of the twist operators, we evaluate it as
\bea
&& \tr_B \r_B^n = \Big( \f{\e}{\ell} \Big)^{2h_\s} \Big\{
[ 1 + \ell^2 b_T \lag T \rag_0 + \ell^4( b_\mA \lag \mA \rag_0 + b_{TT} \lag T \rag_0^2) + O(\ell^6) \big] \nn\\
&& \phantom{\tr_B \r_B^n =}
+ \Big[ \ell^4 \sum_{k=1}^{n-1} q^{2k} d_{TT}^{0k} \f{\lag T|T(0)|0\rag\lag 0|T(0)|T\rag}{\a_T} + O(\ell^5) \Big] \nn\\
&& \phantom{\tr_B \r_B^n =}
+q^{2n}[ 1 + \ell^2 b_T \lag T \rag_T + \ell^4 ( b_\mA \lag \mA \rag_T + b_{TT} \lag T \rag_T^2) + O(\ell^6) \big]
 \Big\} + O(q^3).
\eea
We get the long interval entanglement entropy
\bea
&& S_\lo(\ell) = \f{c}{6}\log\f{\ell}{\e} + \big[ \ell^2 a_T \lag T \rag_0 + \ell^4 a_{TT} \lag T \rag_0^2 + O(\ell^6) \big]\\
&& \phantom{S_\sh(\ell) =}
   + \Big[ -\ell^4 \p_n \Big( \sum_{k=1}^{n-1} q^{2k} d_{TT}^{0k} \Big)\Big|_{n\to1} \f{\lag T|T(0)|0\rag\lag 0|T(0)|T\rag}{\a_T} + O(\ell^5) \Big] \nn\\
&& \phantom{S_\sh(\ell) =}
   + \Big[ 1 + \f{4\pi\b}{L} + \ell^2 a_T ( \lag T \rag_T - \lag T \rag_0) + \ell^4 a_{TT} ( \lag T \rag_T^2 - \lag T \rag_0^2) + O(\ell^6) \Big]q^2 + O(q^3).\nn
\eea
Using (\ref{e72}), (\ref{e75}), (\ref{e73}), we finally get
\bea \label{e78}
&& S_\lo(L-\ell) = \f{c}{6}\log\f{\ell}{\e} + \Big( -\f{\pi^2 c \ell^2}{36L^2} -\f{\pi^4 c \ell^4}{1080L^4} + O(\ell^6) \Big)
                 +\Big[ - \Big( \f{\pi\ell}{L} \Big)^4 \p_n \Big( \sum_{k=1}^{n-1}\f{q^{2k}}{\sin^4\f{\pi k}{n}} \Big)\Big|_{n=1}
                        + O(\ell^5) \Big]\nn\\
&& \phantom{S_\lo(L-\ell) =}
+ \Big( 1 + \f{4\pi\b}{L} + \f{4\pi^2 \ell^2}{3L^2} + \f{4\pi^4 \ell^4}{45L^4} + O(\ell^6) \Big) q^2 + O(q^3).
\eea
This is consistent with (\ref{e68}).

Using the thermal entropy (\ref{e76}), the short interval and long interval entanglement entropies (\ref{e77}), (\ref{e78}), we get the correction to the entanglement plateau
\be
S(L) - S_\lo(L-\ell) + S_\sh(\ell) = \Big( \f{\pi\ell}{L} \Big)^4 \Big[ \f{32q^2}{15c} + \p_n \Big( \sum_{k=1}^{n-1}\f{q^{2k}}{\sin^4\f{\pi k}{n}} \Big)\Big|_{n=1} \Big] + O(q^3,\ell^5).
\ee
The second term in the bracket has been evaluated in the previous appendix.
Finally we find that this is the same as the result (\ref{z12}) in section~\ref{sec3}.

\subsection{Contributions from a nonvacuum module}

We consider the contributions from a holomorphic primary operator $\mX$  to the density matrix (\ref{e79}).
Generally, the OPE of the twist operators can be written as
\bea
&& \s(\ell)\td\s(0) = \Big( \f{\e}{\ell} \Big)^{2h_\s} \Big\{ 1
+ \sum_\mY \Big[ \sum_{j=0}^{n-1} \ell^{h_\mY}d_\mY \mY_j(0) + O(\ell^{h_\mY+1}) \Big] \nn\\
&& \phantom{\s(\ell)\td\s(0) =}
+ \sum_\mY \Big[ \sum_{0\leq j_1 < j_2 \leq n-1} \ell^{2h_\mY}d_{\mY\mY}^{j_1j_2} \mY_{j_1}(0)\mY_{j_2}(0)
                 + O(\ell^{2h_\mY+1},\ell^{3h_\mY}) \Big]
\Big\},
\eea
with the summation of $\mY$ being over $T$ and all holomorphic primary operators.

For the short interval $A$ we have the correction to the partition function
\bea
&& \d_\mX \tr_A\r_A^n = \sum_{k=1}^n \f{q^{k h_\mX}}{\a_\mX^k} \sum_{0 \leq j_1 < \cdots < j_k \leq n-1}
\lag \mX_{j_1}\cdots \mX_{j_k}| \s(\ell)\td\s(0) | \mX_{j_1}\cdots \mX_{j_k} \rag + O(q^{h_\mX+1}) \nn\\
&& \phantom{\d_\mX \tr_A\r_A^n}
= \Big( \f{\e}{\ell} \Big)^{2h_\s} \Big\{ q^{h_\mX} \Big[ n
   + \sum_\mY [ \ell^{h_\mY} b_\mY \lag\mY\rag_\mX + O(\ell^{h_\mY+1}) ]\\
&& \phantom{\d_\mX \tr_A\r_A^n=}
   + \sum_\mY [ \ell^{2h_\mY} b_{\mY\mY}\lag\mY\rag_0(2\lag\mY\rag_\mX-\lag\mY\rag_0) + O(\ell^{2h_\mY+1},\ell^{3h_\mY}) ]
\Big] + O(q^{h_\mX+1},(n-1)^2) \Big\}. \nn
\eea
Then we get the corrections to the short interval entanglement entropy
\bea
&&\hspace{-10mm} \d_\mX S_\sh(\ell) = q^{h_\mX} \Big\{
     \sum_\mY [ \ell^{h_\mY} a_\mY ( \lag\mY\rag_\mX - \lag\mY\rag_0 )  + O(\ell^{h_\mY+1}) ] \nn\\
&&\hspace{-10mm} \phantom{\d_\mX S_\sh(\ell) =}
   + 2 \sum_\mY [ \ell^{2h_\mY} a_{\mY\mY} \lag\mY\rag_0 ( \lag\mY\rag_\mX - \lag\mY\rag_0 )  + O(\ell^{2h_\mY+1},\ell^{3h_\mY}) ]
\Big\} + O(q^{h_\mX+1},q^{2h_\mX}).
\eea

For the long interval $B$ we have corrections to the partition function \cite{Chen:2014ehg}
\bea
&& \d_\mX \tr_B \r_B^n = \sum_{k=1}^n \f{q^{k h_\mX}}{\a_\mX^k} \sum_{0 \leq j_1 < \cdots < j_k \leq n-1}
\lag \mX_{j_1}\cdots \mX_{j_k}| \s(\ell)\td\s(0) | \mX_{j_1+1}\cdots \mX_{j_k+1} \rag + O(q^{h_\mX+1}) \nn\\
&& \phantom{\d_\mX \tr_B \r_B^n} =
   \Big( \f{\e}{\ell} \Big)^{2h_\s} \Big\{
      \Big[ \ell^{2h_\mX} \sum_{k=1}^{n-1}q^{k h_\mX}d_{\mX\mX}^{0k} \f{\lag \mX|\mX(0)|0\rag\lag 0|\mX(0)|\mX\rag}{\a_\mX}
           + O(\ell^{2h_\mX+1})
      \Big] \nn\\
&& \phantom{\d_\mX \tr_B \r_B^n=}
   + q^{n h_\mX} \Big[ 1 + \sum_\mY [ \ell^{h_\mY}b_\mY \lag\mY\rag_\mX + O(\ell^{h_\mY+1})
                       + \sum_\mY [ \ell^{2h_\mY}b_{\mY\mY} \lag\mY\rag_\mX^2 + O(\ell^{2h_\mY+1},\ell^{3h_\mY}) ] \Big]\nn\\
&& \phantom{\d_\mX \tr_B \r_B^n=}
   + O(q^{h_\mX+1}) \Big\},
\eea
from which we get the corrections to the long interval entanglement entropy
\bea
&& \d_\mX S_\lo(L-\ell) =
    \Big[ -\ell^{2 h_\mX} \p_n \Big( \sum_{k=1}^{n-1} q^{k h_\mX} d_{\mX\mX}^{0k} \Big)\Big|_{n\to1} \f{\lag \mX|\mX(0)|0\rag\lag 0|\mX(0)|\mX\rag}{\a_\mX} + O(\ell^{2h_\mX+1}) \Big] \nn\\
&& \phantom{\d_\mX S_\lo(L-\ell) =}
   + q^{h_\mX} \Big\{ 1 + \f{4\pi\b h_\mX}{L}
   + \sum_\mY [ \ell^{h_\mY} a_\mY ( \lag\mY\rag_\mX - \lag\mY\rag_0 ) + O(\ell^{h_\mY+1}) ] \nn\\
&& \phantom{\d_\mX S_\lo(L-\ell) =}
   + \sum_\mY [ \ell^{2h_\mY} a_{\mY\mY} ( \lag\mY\rag_\mX^2 - \lag\mY\rag_0^2 )  + O(\ell^{2h_\mY+1},\ell^{3h_\mY}) ]
+ O(q^{h_\mX+1},q^{2h_\mX}).
\eea

The corrections to the thermal entropy are (\ref{e80}). Then we get the corrections to the entanglement plateau
\bea
&& \hspace{-6mm}
   \d_\mX ( S(L) - S_\lo(L-\ell) + S_\sh(\ell) ) =
    \Big[ \ell^{2 h_\mX} \p_n \Big( \sum_{k=1}^{n-1} q^{k h_\mX} d_{\mX\mX}^{0k} \Big)\Big|_{n\to1} \f{\lag \mX|\mX(0)|0\rag\lag 0|\mX(0)|\mX\rag}{\a_\mX} + O(\ell^{2h_\mX+1}) \Big] \nn\\
&& ~~~~~~~~~
   - q^{h_\mX} \sum_\mY \big[ \ell^{2h_\mY} a_{\mY\mY} ( \lag\mY\rag_\mX - \lag\mY\rag_0 )^2  + O(\ell^{2h_\mY+1},\ell^{3h_\mY}) \big]
 + O(q^{h_\mX+1},q^{2h_\mX}).
\eea
Note that the summation of $\mY$ is over $T$ and all holomorphic primary operators. Using (\ref{e75}), (\ref{e82}), (\ref{e73}), (\ref{e83}), we can show easily that this is exactly the same as the result (\ref{z14}).

\providecommand{\href}[2]{#2}\begingroup\raggedright\endgroup


\end{document}